\documentclass[5p]{elsarticle}

\usepackage{xspace}
\usepackage{verbatim}
\usepackage{cprotect}
\usepackage{amsmath,amssymb}
\usepackage{tabularx}
\usepackage{todonotes}
\usepackage{multirow}
\usepackage{subfigure}
\usepackage{listings}
\usepackage{hyperref}

\newcommand{\argmax}{\operatornamewithlimits{argmax}}

\definecolor{mygray}{rgb}{0.8,0.8,0.8}
\def\culsp{\textsc{culsp}\xspace}

\def\gpu{\textsc{lsp-gpu}\xspace}
\def\lsppy{\textsc{lsp-py}\xspace}
\def\lspc{\textsc{lsp-c}\xspace}
\def\sincos{\texttt{sincos}\xspace}
\def\sincosfintrinsic{\texttt{\_\_sincosf()}\xspace}
\def\sin{\texttt{sin}\xspace}
\def\cos{\texttt{cos}\xspace}

\def\ida{\textit{243Ida}\xspace}

\def\platforma{\textsc{Platform1}\xspace}
\def\platformb{\textsc{Platform2}\xspace}

\newcolumntype{R}{>{\raggedleft\arraybackslash}X}
\newcolumntype{Y}{>{\centering\arraybackslash}X}

\journal{Journal of \LaTeX\ Templates}

\biboptions{authoryear}

\begin{document}

\begin{frontmatter}

\title{Fast Period Searches Using the Lomb-Scargle Algorithm on Graphics Processing Units for Large Datasets and Real-Time Applications}


\author[NAU1]{Michael Gowanlock\corref{mycorrespondingauthor}}
\cortext[mycorrespondingauthor]{Corresponding author}
\ead{michael.gowanlock@nau.edu}
\author[NAU1,NAU2]{Daniel Kramer}
\author[NAU2]{David E. Trilling}
\author[ASU]{Nathaniel R. Butler}
\author[NAU1]{Brian Donnelly}


\address[NAU1]{School of Informatics, Computing, and Cyber Systems, Northern Arizona University, Flagstaff, AZ, 86011, USA}
\address[NAU2]{Department of Astronomy \& Planetary Science, Northern Arizona University, Flagstaff, AZ, 86011, USA}
\address[ASU]{School of Earth \& Space Exploration, Arizona State University, Tempe, AZ, 85287, USA}

\begin{abstract}
Computing the periods of variable objects is well-known to be computationally expensive. Modern astronomical catalogs contain a significant number of observed objects. Therefore, even if the period ranges for particular classes of objects are well-constrained due to expected physical properties, periods must be derived for a tremendous number of objects. In this paper, we propose a GPU-accelerated Lomb-Scargle period finding algorithm that computes periods for single objects or for batches of objects as is necessary in many data processing pipelines. We demonstrate the performance of several optimizations, including comparing the use of shared and global memory GPU kernels and using multiple CUDA streams to copy periodogram data from the GPU to the host. Also, we quantify the difference between 32-bit and 64-bit floating point precision on two classes of GPUs, and show that the performance degradation of using 64-bit over 32-bit is greater on the CPU than a GPU designed for scientific computing.  We find that the GPU algorithm achieves superior performance over the baseline parallel CPU implementation, achieving a speedup of up to 174.53$\times$. The Vera C. Rubin Observatory will carry out the Legacy Survey of Space and Time (LSST). We perform an analysis that shows we can derive the rotation periods of batches of Solar System objects at LSST scale in near real-time, which will be employed in a future LSST event broker. All source code has been made publicly available.
\end{abstract}


\end{frontmatter}


\section{Introduction} \label{sec:intro}
The Lomb-Scargle Periodogram (LSP) algorithm is a search approach used to find the periods of objects observed at uneven time intervals \citep{1976Ap&SS..39..447L,1982ApJ...263..835S}. The algorithm searches a frequency grid and returns a Lomb-Scargle (L-S) power for each frequency, where high L-S powers indicate a potential periodic signal in the time series. 

The na\"{i}ve LSP algorithm has a time complexity of $O(N_t^2)$, where $N_t$ is the number of observations in the time series \citep{2010ApJS..191..247T}, which makes the algorithm computationally expensive. In data processing pipelines that require derived properties of objects, such as periodicity, the LSP algorithm can be a major bottleneck in the pipeline. Consequently, several studies have proposed new algorithms that have reduced the time complexity to $O(N_t\mathrm{log}N_t)$ \citep{1989ApJ...338..277P,2012A&A...545A..50L}, which may come at the expensive of accuracy. For clarity, in this paper, we consider the $O(N_t^2)$ LSP algorithm.

Graphics Processing Units (GPUs) and associated software support have evolved rapidly over the past two decades. Initially, GPUs did not provide application programming interface (API) support for general purpose computing (i.e., non-graphics applications), but APIs such as CUDA \citep{CUDAbook}, OpenCL \citep{munshi2009opencl}, ROCm \citep{sun2018evaluating}, and OpenACC \citep{farber2016parallel} now allow programmers straightforward access to the functionality of modern GPU hardware. Additionally, there are now two major GPU markets, a consumer market, and a scientific computing/artificial intelligence/data center market. Scientific applications may be sensitive to the level of floating point precision offered by the hardware (e.g., 64-bit floating point precision may be required for numerical stability in dynamical simulations). Consumer-grade GPUs typically have most of their resources dedicated to 32-bit floating point arithmetic, whereas the GPUs designed for the scientific computing  market offer hardware dedicated to a wide range of precision levels, from 16-bit to 64-bit floating point values \citep{haidar2018harnessing}. 

\citet{2010ApJS..191..247T} proposed an $O(N_t^2)$ LSP algorithm programmed in CUDA, and this paper builds on that work. In particular, we add additional functionality to the algorithm to allow for a wider range of use cases. Furthermore, we propose several optimizations that reflect recent hardware advances. Beyond optimizations that improve performance in GPU kernels, we summarize the features of the proposed algorithm, \gpu, as follows.

\begin{itemize}
\item Computing a single periodogram for one object or a batch of periodograms for multiple objects.
\item Storage of the input data and computation on both 32-bit and 64-bit floating point data. 
\item Option to return the periodogram to the host for both the single object mode and batch mode.
\item To improve the performance of transferring periodogram data from the GPU to the host, there is an option to transfer the periodogram data in several CUDA streams using small pinned memory staging buffers. Additionally, the data is transferred from the staging buffers into pageable memory in parallel using the CPU.
\item We implement both the standard and generalized LSP algorithms, where the latter includes photmetric error and floats the mean.
\end{itemize}

The paper is organized as follows. Section~\ref{sec:LSPoverview} gives a brief outline of the LSP algorithm. Section~\ref{sec:gpu} presents our GPU-accelerated LSP algorithm, \gpu, describing the proposed optimizations and functionality of our software.  Section~\ref{sec:evaluation} presents the evaluation of our algorithm compared to a parallel CPU reference implementation. Finally, Section~\ref{sec:conclusion} concludes the work and outlines future research avenues.

\section{Lomb-Scargle Periodogram Algorithm}\label{sec:LSPoverview}
In this section, we briefly define the Lomb-Scargle Periodogram and notation that we use throughout the paper. For additional information on LSP, we refer the reader to the original papers \citep{1976Ap&SS..39..447L,1982ApJ...263..835S} and a comprehensive review of the LSP algorithm by \citet{2018ApJS..236...16V}.

Consider a time series with $N_t$ measurements, which are (unevenly) sampled at times $t_j$, where $j=1,\ldots,N_t$. Each measurement (i.e., magnitude) at time $t_j$ is denoted as $X_j$. We assume here that the mean is normalized to zero. 

In this paper, we assume that all summations are from $j=1,\ldots,N_t$, and $\omega=2\pi f$ is the angular frequency. The time delay $\tau$ is defined as follows:

\begin{equation}
\mathrm{tan}~2\omega\tau=\frac{\Sigma_j~\mathrm{sin}~2\omega t_j}{\Sigma_j~\mathrm{cos}~2\omega t_j}.
\end{equation}

The LSP as a function of frequency, $f$, is

\begin{eqnarray}
P_{LS}(f)=\frac{1}{2}\biggl(\frac{[\Sigma_j X_j~\mathrm{cos}~\omega(t_j-\tau)]^2}{\Sigma_j~\mathrm{cos}^2~\omega(t_j-\tau)}+\nonumber\\
\frac{[\Sigma_j X_j~\mathrm{sin}~\omega(t_j-\tau)]^2}{\Sigma_j~\mathrm{sin}^2~\omega(t_j-\tau)}\biggr).
\label{eqn:LSP}\end{eqnarray}

As discussed in \citet{2010ApJS..191..247T}, directly utilizing Equation~\ref{eqn:LSP} in a computer program is inefficient, as it requires two scans over the time series. Alternatively, the equation can be rewritten to use several constants that yield a single scan over the time series \citep[see][for details]{press1992numerical,2010ApJS..191..247T}. This improved algorithm is standard in many implementations. Our GPU and CPU implementations are directly ported from the SciPy implementation that employs the refactored equation\footnote{\url{https://docs.scipy.org/doc/scipy/reference/generated/scipy.signal.lombscargle.html}}. 

Throughout this paper, we assume an evenly spaced frequency grid. We denote $N_f$ as the number of frequencies searched within the frequency range $[f_{min}, f_{max})$. Therefore, the frequency spacing is given by $\Delta f=(f_{max}-f_{min})N_f^{-1}$.

The values of $N_f$, $f_{min}$, and $f_{max}$ will impact the quality of the result. First, domain knowledge is required to select expected frequency limits. It may be reasonable to select $f_{min}\approx0$ for most applications \citep{2018ApJS..236...16V}, but $f_{max}$ will need to be determined based on the expected physical characteristics of an object. Additionally, if $\Delta f$ is too large then the algorithm may miss the peaks in the periodogram. We do not provide a method of selecting these parameters, as they depend on several assumptions about the object and oversampling rate, but we refer the reader to \citet{2010ApJS..191..247T} and  \citet{2018ApJS..236...16V} for a broader discussion of input parameter selection.

\section{GPU-Accelerated Lomb-Scargle}\label{sec:gpu}
In this section, we present an overview of GPUs and the code, the two supported modes for computing periodograms, a data transfer optimization to reduce the overhead of copying the periodogram data from the GPU to main memory, and kernel designs that use global and shared memory. Note that we use CUDA terminology throughout this paper. 

\subsection{Graphics Processing Units and the Lomb-Scargle Periodogram}
GPU architectures contain different amounts of hardware dedicated to certain types of arithmetic. Some recent generations of the Nvidia Tesla and Quadro GPUs contain many resources dedicated to 64-bit floating point operations, whereas others contain few resources dedicated to supporting these operations. For example, the Nvidia RTX Turing TU102 GPU contains only 1/32nd of the hardware dedicated to 64-bit floats (FP64) as 32-bit floats (FP32)\footnote{\url{https://www.nvidia.com/content/dam/en-zz/Solutions/design-visualization/technologies/turing-architecture/NVIDIA-Turing-Architecture-Whitepaper.pdf}}. In contrast, this ratio is 1/2 on the Nvidia Ampere A100 architecture\footnote{\url{https://www.nvidia.com/content/dam/en-zz/Solutions/Data-Center/nvidia-ampere-architecture-whitepaper.pdf}}. Consequently, if an application requires the precision provided by FP64 arithmetic, then GPUs for scientific computing and data centers, such as Pascal (P100), Volta (V100), and Ampere (A100) are preferable to other GPUs that may not provide high FP64 throughput.

Modern GPUs hide high memory access latency through rapid context switching in hardware. The GPU has a large number of registers that accommodate many threads on a single streaming multiprocessor (SM). The large number of registers (or context) are used to allow one set of threads to stall for memory while another set of threads execute on the SM. This design enables the SMs to be utilized despite threads stalling for memory. One drawback of FP64 is that they require more registers than FP32. This increases register pressure, which may limit the number of threads that can be active at any given time. One primary design goal of \gpu is to support both FP32 and FP64 such that users can use a variety of GPU hardware to compute the LSP at varying degrees of accuracy. Since LSP computes many mathematical operations, and uses a significant number of registers, using FP64 is significantly more expensive than FP32.

\subsection{Overview of Program Elements Common to Both Modes}
Our code contains a mode to compute the Lomb-Scargle periodogram for a single object. In addition, we also include a mode to simultaneously compute the LSP for several objects in parallel. In this section, we give a brief overview of the elements of the code that are common to both modes. Note that throughout this section, we refer to the standard LSP algorithm and not the generalized algorithm that floats the mean and incorporates photometric error. We will discuss the generalized algorithm in Section~\ref{sec:generalized_LSP}.

The \gpu code has been modified from the SciPy LSP algorithm, which has a time complexity of $O(N_fN_t)$. We made few modifications to the code.
We leverage the \sincos function in the CUDA math API which simultaneously computes both the sine and cosine of a value. This eliminates performing independent sine and cosine calculations. LSP requires computing \sincos twice in the kernel, which we found to outperform two separate calls to \sin and \cos.

\subsection{Single Periodogram Mode}\label{sec:gpusinglepgram}
The single periodogram mode is used in the case where the user wants to compute a large input time series and/or a large number of frequencies.

The user provides a minimum and maximum frequency to examine in the range $[f_{min}, f_{max})$, and the number of frequencies to compute, $N_f$. These input parameters yield the frequency spacing $\Delta f$. To compute the periodogram, we parallelize the computation across frequencies. Using $N_f$ total computed frequencies, we launch $N_f$ threads, where each thread is assigned a single frequency to compute.  Each thread stores the computed L-S power for its assigned frequency in global memory. Then, we execute another kernel that performs $\argmax_x(pgram)$ to find the index, $x$, in the periodogram that contains the maximum L-S power. The  index $x$ is returned to the host, and then converted into the period as follows: $p=2\pi(f_{min}+x\Delta f)^{-1}$.    The $\argmax$ operation is computed using the \texttt{thrust::max\_element} function in the Thrust library \citep{bell2012thrust} which performs a parallel reduction to find $x$.  

We have included the capability to allow the user to determine whether they would like to return the periodogram to the host. If the user prefers, they can simply leave the periodogram in global memory on the GPU.  As we will discuss, returning the periodogram to the host takes non-negligible time, so the user may wish to discard the periodogram if they only want the period corresponding to the frequency with the greatest L-S power. 

The algorithm requires global memory space, including $2N_t$ to store the time series measurements for the sampled times and  magnitudes, in addition to the resulting periodogram of size $N_f$. Therefore, the space complexity is $O(N_t+N_f)$.

\subsection{Batch Periodogram Mode}
The batch periodogram mode is used to concurrently compute periods for multiple objects and is the primary motivation for this work. The Solar System Notification and Alert Processing System (SNAPS) is a planned Vera C. Rubin Observatory Legacy Survey of Space and Time (LSST) \citep{2019ApJ...873..111I} event broker that will send alerts to the astronomy community regarding Solar System objects. LSST has a visit exposure time of 30 seconds \citep{2009arXiv0912.0201L}; therefore, the event broker will receive alerts for $\sim$1,000 Solar System objects every 30~s. Consequently, it is imperative that the period finding algorithm be executed as fast as possible to leave sufficient time to carry out outlier detection activities that rely on the rotation period feature.    

In the batched periodogram mode, we use the same frequency grid for all objects, as we assume that the objects have similar physical properties that would limit the frequency ranges and appropriate value of $\Delta f$. We denote the number of objects in the batch as $N_o$. During the first few years of LSST, an appropriate period range will span $\sim$1 to $\gtrsim$2,000 hours because we will not know an object's a priori rotation period~\citep[see][which shows that asteroids can have rotation periods in a very large range]{warner2009asteroid}. Thus, to ensure that we do not exclude those objects with long rotation periods, we will use a large frequency range.

One option to compute the LSP for multiple objects is to execute the single periodogram kernel described in Section~\ref{sec:gpusinglepgram} for each object. However, there are several drawbacks to this approach. First, if we assume that the LSP would need to be computed for a batch of $\sim$1,000 objects, then this would require $\sim$1,000 kernel invocations, which would incur non-negligible invocation overhead. Second, executing several small kernels requires the GPU's hardware scheduler to perform extra work to allow the kernels to concurrently use the GPU's resources. These drawbacks can be avoided by launching a single kernel that computes the results for all objects, and we elect to use this approach.

The algorithm largely differs from the single periodogram kernel in the way that it is parallelized. Given the time series of multiple objects, we parallelize the L-S algorithm by assigning a single CUDA thread block to compute the periodogram for each object, where each thread may compute multiple frequencies in the periodogram for a given object. This design also allows threads computing the same object within a thread block to share information in shared memory. Shared memory is a form of scratchpad memory in computer architecture that is used for temporary calculations that is faster to access than other memory locations. On the GPU, each SM has a small amount of shared memory that resides on-chip and is therefore faster to access than global memory which is shared by all SMs and exists off-chip.  

As discussed for the single periodogram mode, we have included the option of returning the periodogram to the host at the end of the computation. If we do not want to store the periodogram for each object, instead of storing the periodogram in global memory, as was the case for the single periodogram mode, we have each thread in each CUDA block keep track of the highest power it has computed and the frequency index at which it was found. Each thread stores these pairs in shared memory, and after the powers have been computed for all frequencies, we perform a parallel reduction in shared memory to find the maximum power and frequency index which yields the period. After the period has been computed in shared memory, one thread writes this information to global memory. After the kernel finishes executing, we transfer the periods of all objects back to the host. 

An alternative to selecting the period with the greatest L-S power would be to return $k$ periods corresponding to the top-$k$ powers. We have not included this functionality, as users wanting to employ our algorithm for offline (manual) data analysis tasks are likely to perform the analysis by examining the periodograms of all objects in the input dataset. In the online processing case, such as using the algorithm in an event broker, there will not be sufficient time to manually inspect the periodograms, so we do not capture the top-$k$ periods.

If we return the periodogram to the host, then we simply transfer one array containing the periodograms for all objects and then find the periods on the CPU instead of the GPU. This eliminates the need to have each thread keep track of the maximum power it has found and the parallel reduction step in shared memory described above is unnecessary. 

The use of shared memory in this mode of operation also allows us to reduce register pressure. We can store information common to all threads in the block in shared memory instead of using registers. Such information includes the minimum and maximum values in the data arrays that correspond to the object, $\Delta f$, and the offset for writing the periodogram to global memory.

When returning the periodogram to the host, the space required on the GPU is $2N_t+N_oN_f$, yielding a space complexity of $O(N_t+N_oN_f)$. When the periodogram does not need to be returned to the host, the space complexity is $O(N_t+N_o)$.    

\subsection{Transferring the Periodogram to the Host}\label{sec:data_transfer_host}
It is well documented that PCIe data transfers are a bottlebeck in GPU computing \citep{Fujii2013,van2014performance,GOWANLOCK2019}. In the case of the L-S algorithm, transferring periodogram(s) from the GPU back to the host requires a non-negligible amount of time. To reduce this bottleneck, we employ the methods in \citet{GOWANLOCK2019} that reduce the overhead of performing host/device data transfers. We give a brief overview of the data transfer method here, but refer the interested reader to \citet{GOWANLOCK2019} for more detail. In short, when performing a call to \texttt{cudaMemcpy} to transfer data from the GPU back to the host, the driver must create a temporary pinned memory buffer which is required so that the GPU's direct memory access (DMA) engine can copy data directly to a memory location that is unable to swap to disk (i.e., it is pinned and not pageable). There is overhead in this process, and in the case of the LSP algorithm, it is preferable to allocate a pinned memory staging buffer to incrementally perform the data transfer from the GPU to the host, which increases the data transfer rate over PCIe. We use three CUDA streams to transfer periodogram data on the GPU into pinned memory staging areas, where each stream is assigned an 8 MiB pinned memory buffer. Each data transfer that uses pinned memory calls the \texttt{cudaMemcpyAsync} function. Then the data is copied from pinned memory on the host into its final location in pageable memory. The memory copy from pinned memory to pageable memory is performed in parallel by multiple CPU threads to saturate main memory bandwidth. We compare data transfer approaches in the experimental evaluation.

\subsection{Using Shared Memory in the Kernel}\label{sec:SM}
The GPU-accelerated LSP algorithm of \citet{2010ApJS..191..247T} proposed tiling the computation using shared memory. Since on-chip shared memory is faster than off-chip global memory, and because each selected frequency must be compared to all elements in the input dataset (observation times and magnitudes), each thread in a block can read one data element from global memory and store it in shared memory. Then, each thread computes on all data elemenets in shared memory. The steps of having all threads storing one element in shared memory and then computing on the data and repeating the process until all input data elements have been computed is a programming pattern used in many GPU algorithms \citep{kirk2016programming}.

For both the single object and batched modes, we create global memory and shared memory kernels so that we can assess potential performance gains of the shared memory optimization. The global memory kernel does not perform the intermediate step of reading the input data into shared memory; instead, a thread directly reads all data elements from global memory.

\subsection{Overview of the Code}
We  outline the code for the single object mode that uses global memory, FP64 values, and returns the periodogram to the host. Since there are multiple modes and combinations of optimizations, for brevity, we elect to only illustrate a single configuration. The interested reader can inspect the publicly available code for further details of the other modes and optimizations.

Listing~\ref{lst:kernel} presents the CUDA kernel. As discussed above, the code is directly ported from SciPy and a few modifications were made. For the reader unfamiliar with GPU programming, the code largely varies from the sequential LSP algorithm through the creation of $N_f$ GPU threads, where each thread computes a single frequency. This means that the outer loop that exists in the sequential CPU LSP algorithm that iterates over frequencies is removed, as each thread is responsible for a single loop iteration. The kernel takes as input the array of times, \texttt{x}, the array of magnitudes, \texttt{y}, the values of $f_{min}$, $f_{max}$, $N_t$, and $N_f$, and returns the periodogram, $\texttt{pgram}$. 

The thread id, \texttt{tid}, is a value enumerated from $0,\ldots,N_f-1$ (line~\ref{alg:tid}). This allows each thread to be assigned a single frequency to compute. Lines~\ref{alg:sm1}--\ref{alg:sm2} compute $\Delta f$ and store it in shared memory to be used by all of the threads in the block. While $\Delta f$ could be stored in registers, since it is constant and shared by all threads, we store it in shared memory. Note the call to \texttt{\_\_syncthreads()} which synchronizes the threads in the block to ensure that the value has been set by thread 0 in the block before any other threads are able to compute beyond line~\ref{alg:sm2}.   Lines~\ref{alg:iftid}--\ref{alg:enddataloop} loop over all of the data elements, where only threads with  \texttt{tid} $<N_f$ participate in the computation. If the total number of threads launched is not equal to $N_f$, then a number of leftover threads will be created that cannot perform any computation, where the maximum number of leftover threads is \texttt{blockDim.x}-1.      Lines~\ref{alg:tau}--\ref{alg:pgram} compute the $\tau$ terms and the power for the frequency which is stored in \texttt{pgram} on line~\ref{alg:pgram}. After the kernel finishes executing, the periodogram is returned to the host.

\begin{footnotesize}
\begin{lstlisting}[language=C, basicstyle=\footnotesize, floatplacement=tbp, caption=Listing of the global memory CUDA kernel. \label{lst:kernel}]
__global__ void LSPOneObj(double * x, double * y, 
const double f_min, const double f_max, 
const unsigned int N_t, const unsigned int N_f, 
double * pgram)
{
(*@\label{alg:tid}@*)  unsigned int tid=(blockIdx.x*blockDim.x)+threadIdx.x; 
  
  //All threads use deltaf in SM
(*@\label{alg:sm1}@*)  __shared__ double deltaf; 
  if (threadIdx.x==0){
    deltaf=(f_max-f_min)/(N_f*1.0);  
  }
(*@\label{alg:sm2}@*)  __syncthreads();

 double c, s, tau, c_tau, s_tau, c_tau2, s_tau2, cs_tau;

(*@\label{alg:iftid}@*)  if (tid<N_f){
    double freqToTest=f_min+(deltaf*tid);
    double xc = 0.0, xs = 0.0, cc = 0.0; 
    double ss = 0.0, cs = 0.0;

    #pragma unroll
    for (int j=0; j<N_t; j++){ 
        sincos(freqToTest * x[j], &s, &c);
        xc += y[j] * c;
        xs += y[j] * s;
        cc += c * c;
        ss += s * s;
        cs += c * s;
    }(*@\label{alg:enddataloop}@*)
    
    tau = atan2(2.0 * cs, cc - ss) / (2.0 * freqToTest); (*@\label{alg:tau}@*)
    sincos(freqToTest * tau, &s_tau, &c_tau);        
    c_tau2 = c_tau * c_tau;
    s_tau2 = s_tau * s_tau;
    cs_tau = 2.0 * c_tau * s_tau;
    
    double f1 = (c_tau * xc + s_tau * xs);
    double f2 = (c_tau * xs - s_tau * xc);
    double d1 = (f1*f1);
    double d2 = (c_tau2 * cc + cs_tau * cs + s_tau2 * ss);
    double d4 = (f2*f2);
    double d5 = (c_tau2 * ss - cs_tau * cs + s_tau2 * cc);

    pgram[tid]=0.5 * ((d1/d2)+(d4/d5)); (*@\label{alg:pgram}@*)
  } //end the if statement tid<N_f
} //end of GPU kernel

\end{lstlisting}
\end{footnotesize}

\section{Experimental Evaluation}\label{sec:evaluation}

\subsection{Experimental Methodology}
The \gpu host code  is written in C and all GPU code is written in CUDA. C programs are compiled with the O3 optimization flag. All time measurements are averaged over three trials, but we exclude the time to normalize the mean amplitude to zero and read the dataset from disk. We use 512 threads per block to execute the \gpu kernels. 

Experiments are conducted on the platforms outlined in Table~\ref{tab:platforms}. The GP100 in \platforma is a GPU designed for data centers and scientific computing and as such has significant resources dedicated to FP64 arithmetic to accommodate applications sensitive to floating point error. In contrast, the TitanX GPU in \platformb is a consumer-grade GPU which has fewer resources dedicated to FP64 operations. Both are Pascal generation GPUs, and are representative of hardware designed for the two major GPU markets.

In the evaluation, we refer to two performance metrics that we clarify here. The speedup is defined as the ratio $T_1/T_2$, where $T_1$ and $T_2$ are two response times. $T_2$ is typically the measurement taken by the optimized algorithm and $T_1$ is a baseline (here, $T_2$ is often the GPU time and $T_1$ is the CPU time).    The parallel efficiency is $(T_s/T_t)t^{-1}$, where $T_s$ and $T_t$  are the sequential and parallel response times, and $t$ is the number of threads/CPU cores used to execute $T_t$. This metric gives an indication of how well the parallel implementation is able to utilize the $t$ CPU cores. Parallel efficiency in this context is not applicable to the GPU.

\begin{table*}[!t]
\centering
\footnotesize
\caption{Details of the platforms used in the experimental evaluation.}\label{tab:platforms}
\begin{tabular}{lllll|llll} 
\hline
& \multicolumn{4}{c|}{\textbf{CPU}} &  \multicolumn{4}{c}{\textbf{GPU}}\\
 Platform& Model & Cores (Total) & Clock & Memory &Model & Cores & Memory & Software\\
\hline
\multicolumn{1}{c}{\platforma} & $2 \times$E5-2620 v4& $2 \times$8 (16)& 2.1 GHz& 128 GiB& Quadro GP100 & 3584 & 16 GiB & CUDA 9\\
\multicolumn{1}{c}{\platformb} & $2 \times$E5-2683 v4& $2 \times$16 (32)& 2.1 GHz& 256 GiB& TitanX & 3584 & 12 GiB & CUDA 10.1\\
\hline
\end{tabular}
\end{table*}

\subsection{Datasets}
\begin{figure}[!t]
\centering
  \includegraphics[width=1.0\columnwidth]{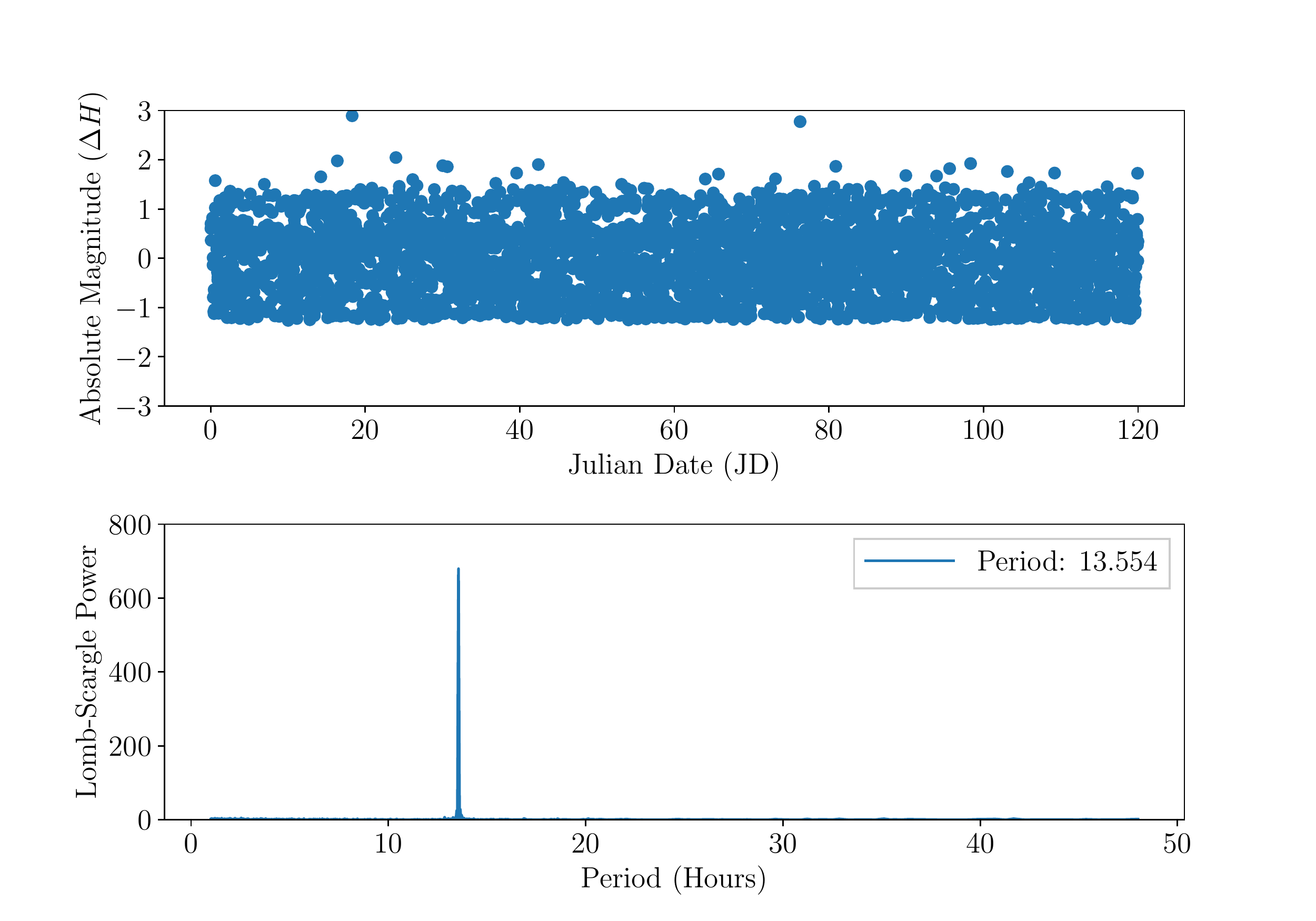}
	\caption{Properties of the dataset used when computing a single periodogram. Upper panel: Mean-subtracted absolute magnitudes of a synthetic object with $N_t=3,554$ measurements. Lower panel: Non-normalized Lomb-Scargle periodogram using $N_f=10^5$, which correctly detects the rotation period of 13.55~h.}
   \label{fig:singleobj}
\end{figure}

We evaluate the L-S algorithm on both the single object and batch processing modes. A suitable dataset for the single object scheme only requires one time series, whereas the batch processing scheme requires input data for multiple objects. For our single object dataset, we generate a time series for a synthetic object with $N_t=3,554$ measurements sampled at uneven time periods. Figure~\ref{fig:singleobj} (top panel) plots the mean-subtracted absolute magnitude ($\Delta H$) time series, and the lower panel shows the L-S periodogram, which correctly finds the rotation period of 13.55~h. We use a time series that has a substantial number of measurements, as it represents the use case where a large time series must be examined. For the single object mode, we do not consider the case where we need to process a small time series, as it is likely that this task can be completed within a reasonable amount of time without GPU acceleration. When we compute the LSP for this object, we use the following frequency ranges $[f_{min}, f_{max})=[3.142, 150.796)$ which correspond to light curve periods of 1--48~h (or rotation periods of 2--96~h), which are typical periods of main belt asteroids in the Solar System.

\begin{figure}[!t]
\centering
\subfigure[]{
  \includegraphics[width=1.0\columnwidth]{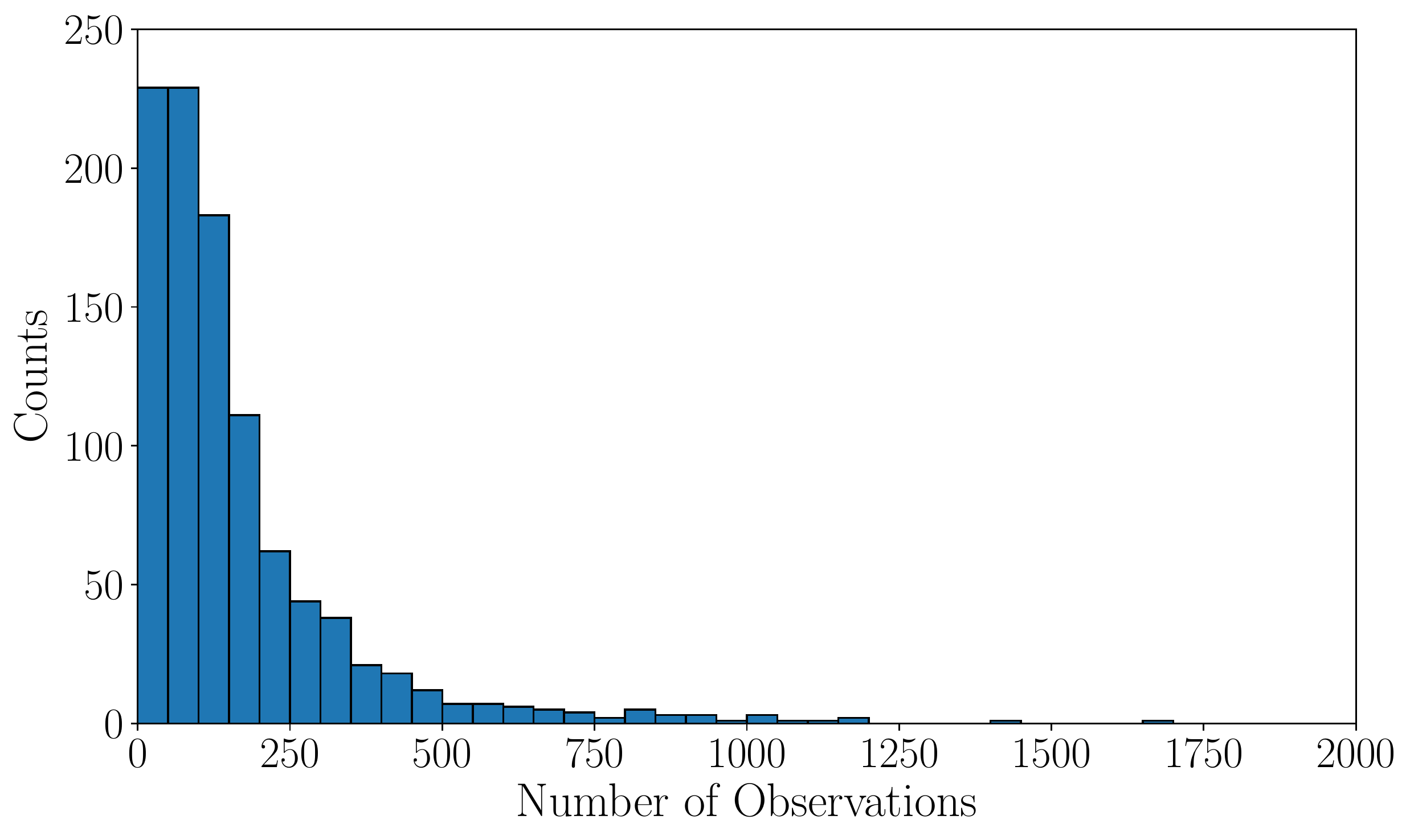}
  }
  \subfigure[]{
  \includegraphics[width=1.0\columnwidth]{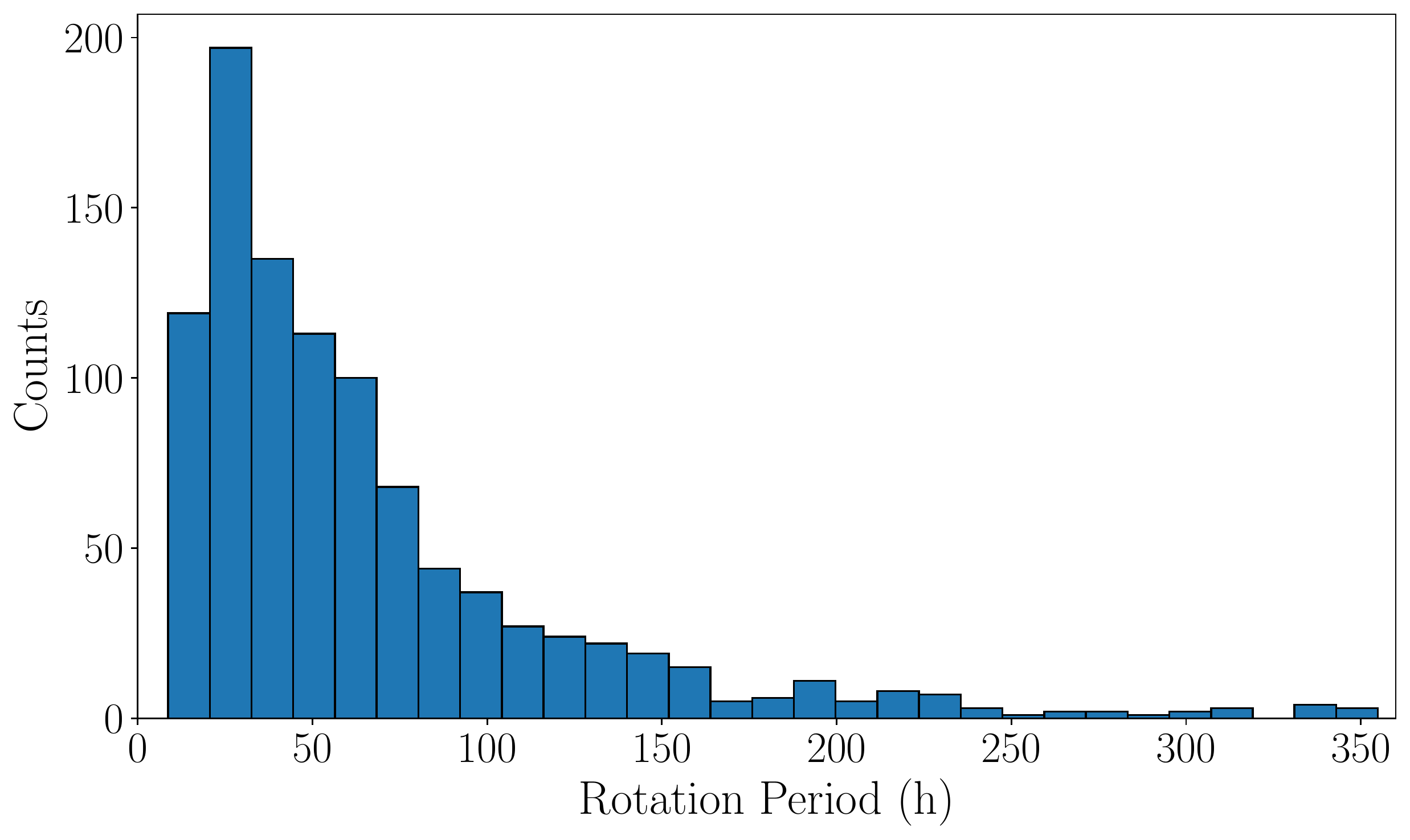}
  }
	\caption{(a) Histogram of the log-normal distribution of the number of observations for the synthetic asteroid population. (b) The rotation period solutions for all objects in the dataset.}
   \label{fig:synthetic_dataset_properties}
\end{figure}

We create a dataset of $\approx1,000$ synthetic asteroids to evaluate the batch processing scheme. Using Zwicky Transient Facility (ZTF) as a pathfinder for LSST, a synthetic observational record is produced –- with ZTF-like cadences and ZTF-like photometric errors.  It is important to capture objects with varying numbers of observations, as this may impact the performance of the GPU algorithm where each object is computed by single CUDA block. Figure~\ref{fig:synthetic_dataset_properties}(a) shows the log-normal distribution of observations for the objects in the dataset. The distribution of the number of observations per object is similar to the number we expect for the LSST towards the end of its lifetime. This allows us to assess our algorithm at LSST scale when deriving periods within the 30~s visit exposure time. When we evaluate the algorithms using batch processing mode, we use the following frequency ranges $[f_{min}, f_{max})=[1.005, 150.796)$ which correspond to rotation periods of $2-300$~h. Figure~\ref{fig:synthetic_dataset_properties}(b) shows the real rotation periods for the synthetic asteroid population.

\begin{figure}[!t]
\centering
\subfigure[]{
  \includegraphics[width=1.0\columnwidth]{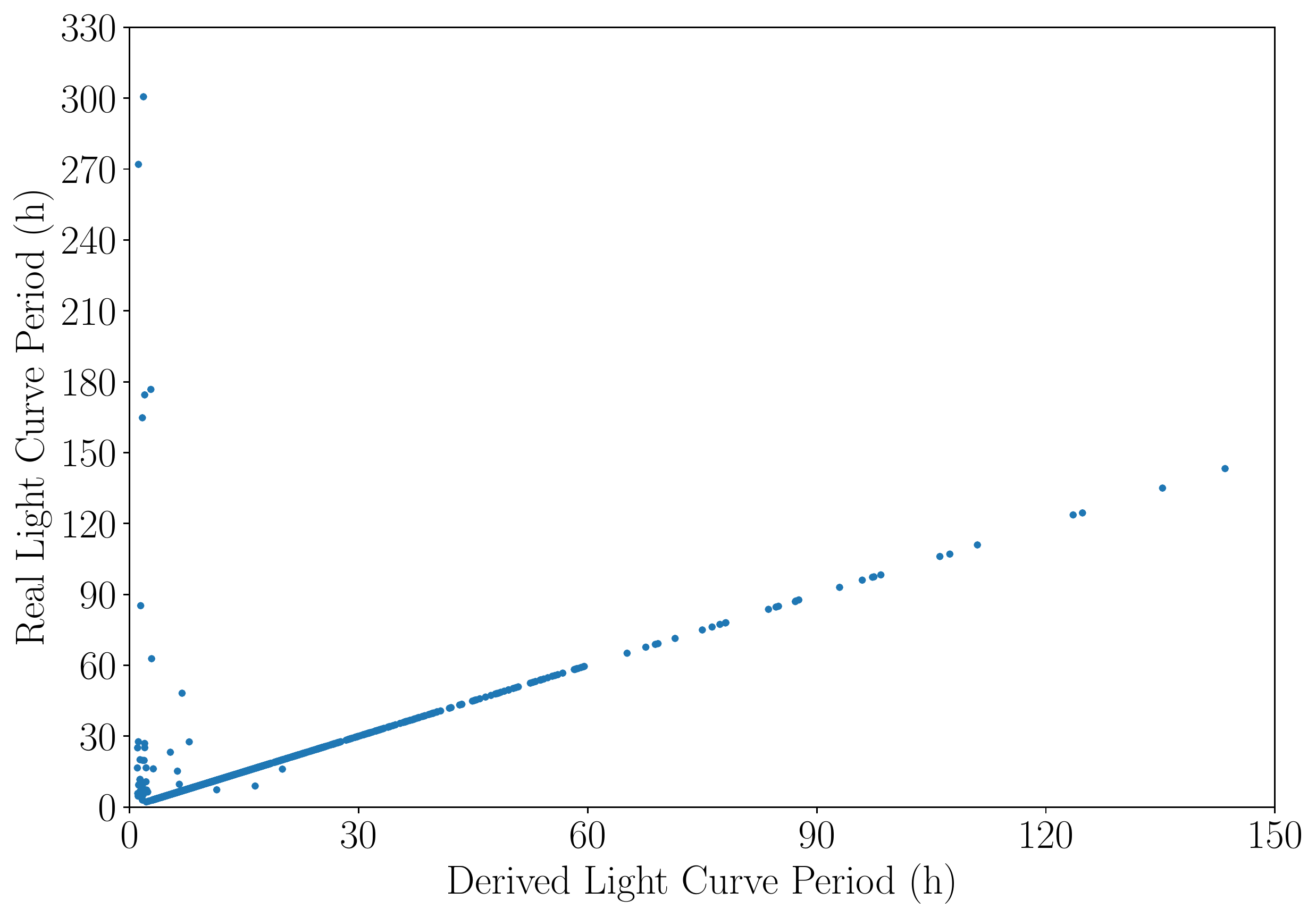}
  }
  \subfigure[]{
  \includegraphics[width=1.0\columnwidth]{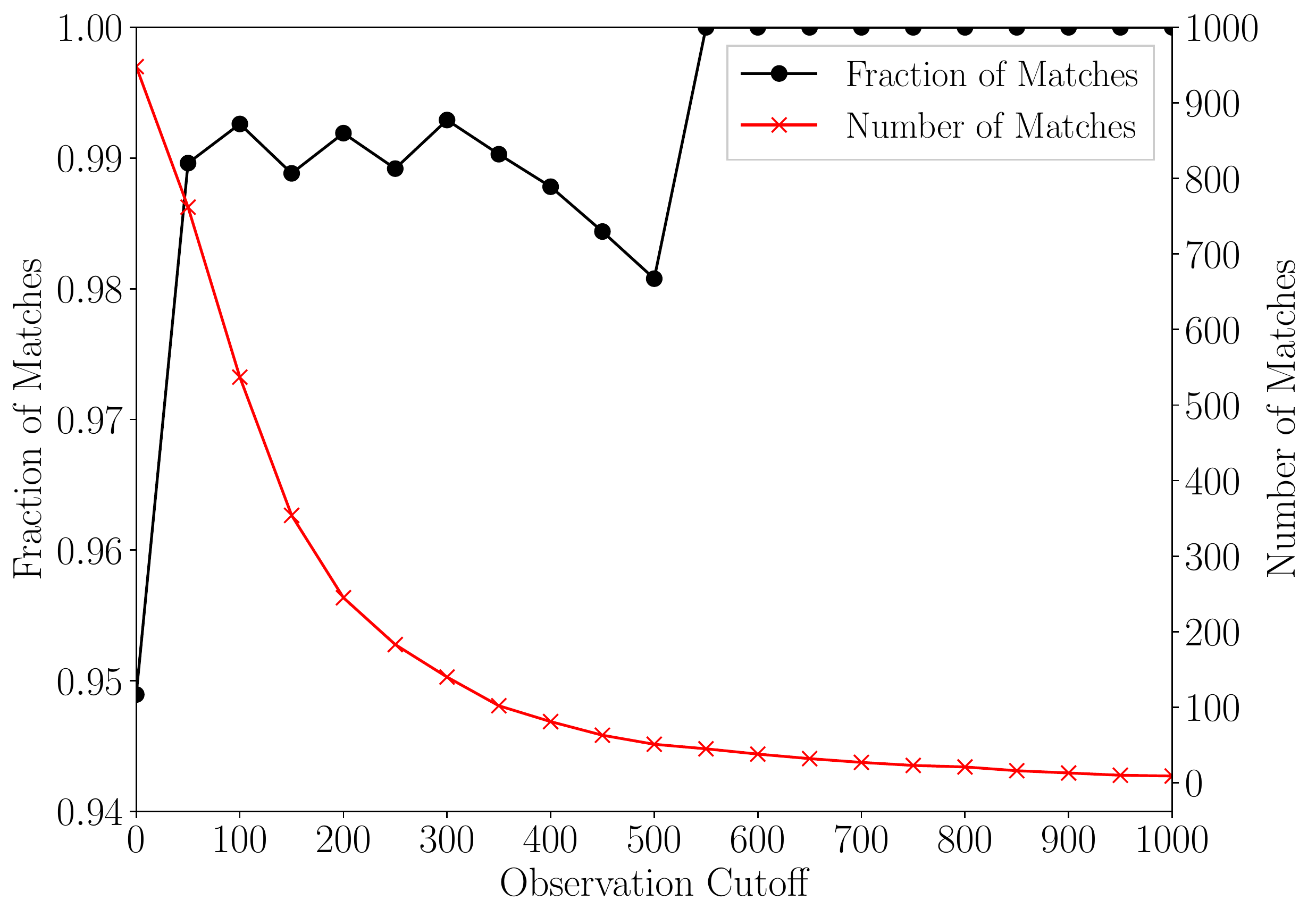}
  }
	\caption{(a) Comparison of the derived vs. assigned periods of the synthetic population of asteroids. The diagonal line indicates that we are able to derive the correct periods to within 0.1~h for a large majority of the population. (b) The fraction of matches (left vertical axis) as a function of the observation cutoff, where each object requires having at least the number of observations shown on the vertical axis. An observation cutoff of 0 includes the entire dataset. The right vertical axis shows the number of objects that match as a function of the observation cutoff. This shows that many of the incorrect derived periods are due to those objects with few observations. The sawtooth pattern at $\geq$200 observations and decrease in the fraction of matches at 500 observations is an artifact of small number statistics.}
   \label{fig:synthetic_dataset_accuracy}
\end{figure}

Figure~\ref{fig:synthetic_dataset_accuracy}(a) plots the real period assigned to our population of synthetic asteroids as a function of the derived periods using the LSP algorithm. We use the maximum  L-S power to determine the period of each object in the dataset, and we use a frequency grid with $N_f=200,000$ frequencies per object. In Figure~\ref{fig:synthetic_dataset_accuracy}(a), a diagonal line indicates general agreement between the derived and real periods.  We find the correct periods for 94.9\% of the objects to within 0.1~h of the real period. Note that a few of the objects have light curve periods $>$150~h, so we were unable to correctly derive those periods. The algorithm does not recover all of the periods due to some objects having too few observations. To demonstrate this, Figure~\ref{fig:synthetic_dataset_accuracy}(b) plots the fraction of matches as a function of the observation cutoff, where an object must have at least the number of observations on the horizontal axes to be included in the sample. An observation cutoff of 0 indicates that all objects in the dataset are included, yielding a 94.9\% agreement (as shown in Figure~\ref{fig:synthetic_dataset_accuracy}(a)). By excluding those objects with $<50$ observations, our agreement between the real and derived periods is 99\%. If we use a cutoff of 550 observations then we achieve a perfect agreement, but only 45 objects are included in the sample. This demonstrates that the source of error in Figure~\ref{fig:synthetic_dataset_accuracy}(a) is due to not removing objects from the dataset that have too few observations.


\subsection{Selection of the Frequency Grid}

The experimental evaluation necessitates examining algorithmic performance across varying values of $N_f$ to understand how the algorithm performs as a function of this parameter. However, some of these values will be too small and undersample the frequency space which may miss the peaks in the periodogram. Here, we outline reasonable values of $N_f$ for each dataset based on the formulation in \citet{2011ApJ...733...10R}.  For a given object, we select $[f_{min}, f_{max})$ based on the science case (the expected range of a periodic signal in the data). We compute the observing window for an object, which is the duration of time between its first and last observation, and denote it as $T^{max}=|t_1-t_{N_t}|$. Then we select $\Delta f=(0.2\pi)/T^{max}$. In the case of computing a batch of objects, $T^{max}$ is the maximum observing window of all objects in the dataset. We then compute $N_f=(f_{max}-f_{min})/\Delta f$. Table~\ref{tab:reasonable_nf} summarizes the frequency ranges for the single object time series and the synthetic population of asteroids used for assessing the performance of batch mode. We ensure that in all experiments that examine the performance as a function of $N_f$, we select a range of values for $N_f$ such that at the very least, we bracket the practical value of $N_f$ outlined in Table~\ref{tab:reasonable_nf}.

\begin{table}[!t]
\centering
\begin{footnotesize}
\caption{Practical values of $N_f$ for the datasets in this paper. Frequency ranges are selected based on scientific objective.} 
\label{tab:reasonable_nf}
\begin{tabularx}{\columnwidth}{|X|r|r|r|} \hline
Dataset&$[f_{min}, f_{max})$ day$^{-1}$& $\Delta f$ & $N_f$\\\hline
Single Object  &$[3.142, 150.796)$&$5.237\times10^{-3}$&28,194\\
Batch &$[1.005, 150.796)$&$5.236\times10^{-3}$&28,608\\
\hline
\end{tabularx}
\end{footnotesize}
\end{table}

\subsection{Accurate Period Finding Demonstration}
\begin{figure}[!t]
\centering
  \includegraphics[width=1.0\columnwidth]{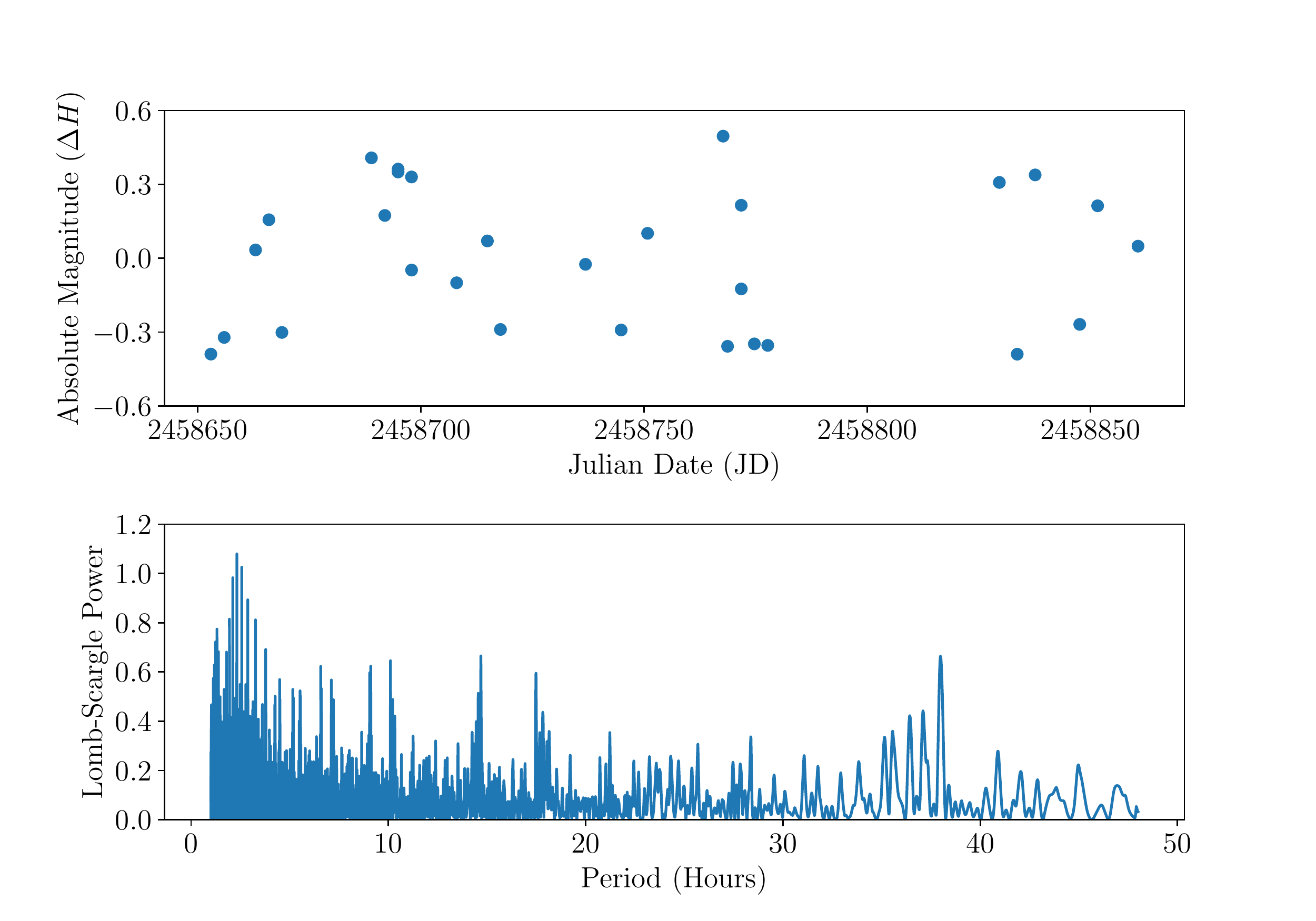}
	\caption{Upper panel: Mean-subtracted absolute magnitude measurements ($\Delta H$) of \ida. Lower panel: Non-normalized Lomb-Scargle periodogram using $N_f=10^5$, which correctly detects the half rotation period of 2.32~h.}
   \label{fig:ida}
\end{figure}

Before we begin evaluating our algorithms, we demonstrate that our LSP algorithm can find a period on a real-world dataset. We select the main belt asteroid 243 Ida that has a known rotation period of 4.63~h \citep{vokrouhlicky2003vector}. We have ingested data from the ZTF public survey \citep{2019PASP..131a8002B}, and use the 28 data points provided by the \texttt{ZTF r} filter. We use the following frequency ranges $[f_{min}, f_{max})=[3.142, 150.796)$ which correspond to rotation periods of $2-96$~h. Figure~\ref{fig:ida} plots $\Delta H$ as a function of time, where the lower panel shows the L-S periodogram. Our algorithm, \gpu, is able to correctly detect the light curve period of 2.32~h, which is half of the rotation period. Due to the geometry of an asteroid, a full rotation will yield two periodic signals. However, L-S will yield a single periodic signal when detecting the period in a time series. Therefore, L-S is expected to produce a light curve with a period that needs to be  doubled to obtain the full rotation period.      

\subsection{Reference Implementations}
We use two implementations for comparison to \gpu. As a sanity check for output accuracy and as a sequential baseline, we use the LSP algorithm in SciPy, which we denote as \lsppy.

Since Python is not easily amenable to parallel execution, and to ensure that we compile with consistent optimizations, we port the algorithm to C and parallelize it with OpenMP and denote this implementation as \lspc. The source code that we port to C is available here\footnote{\url{https://github.com/scipy/scipy/blob/v1.4.1/scipy/signal/_spectral.pyx}}. Since the code from the SciPy implementation is written in Cython, it is straightforward to port to C and only minor modifications were required. 

Regarding the batch mode configuration, we parallelize \lspc on a per-object basis, where each thread computes an entire L-S periodogram for an object. We found that this strategy was more efficient than paralleling within a single object across frequencies. This is because for each object, the threads must be forked and joined, and this overhead is non-negligible. In contrast, when parallelizing on a per-object basis, we only need to create/fork the threads and join them once. Because each object in the batch will not have the same value of $N_t$, we use the dynamic scheduling option in OpenMP, which yields low load imbalance across threads at the end of the computation.

Regarding the single object configuration, we parallelize \lspc across frequencies, where each thread is assigned $N_f/t$ of the loop iterations, where $t$ is the number of threads\footnote{For illustrative purposes, and without loss of generality, we assume $t$ evenly divides $N_f$.}. Since the workload is identical across iterations, we use static scheduling in this parallel OpenMP loop. 

The sequential LSP algorithm iterates over frequencies, which means that the calculation of \sin and \cos can be simply computed using the $\Delta f$ between frequency $n$ and $n+1$, thus eliminating costly \sin and \cos calculations (i.e., the \sincos at iteration $n+1$ in the loop in Listing~\ref{lst:kernel} could be expressed as the difference between the \sincos at iteration $n$). Unfortunately, this optimization eliminates the possibility of parallelization, as it introduces inter-iteration dependencies. Therefore, we do not employ this optimization in \lspc, as we would not be able to execute the algorithm in parallel.

\subsection{Scalability of \lspc}
\begin{figure}[!t]
\centering
  \includegraphics[width=0.45\textwidth]{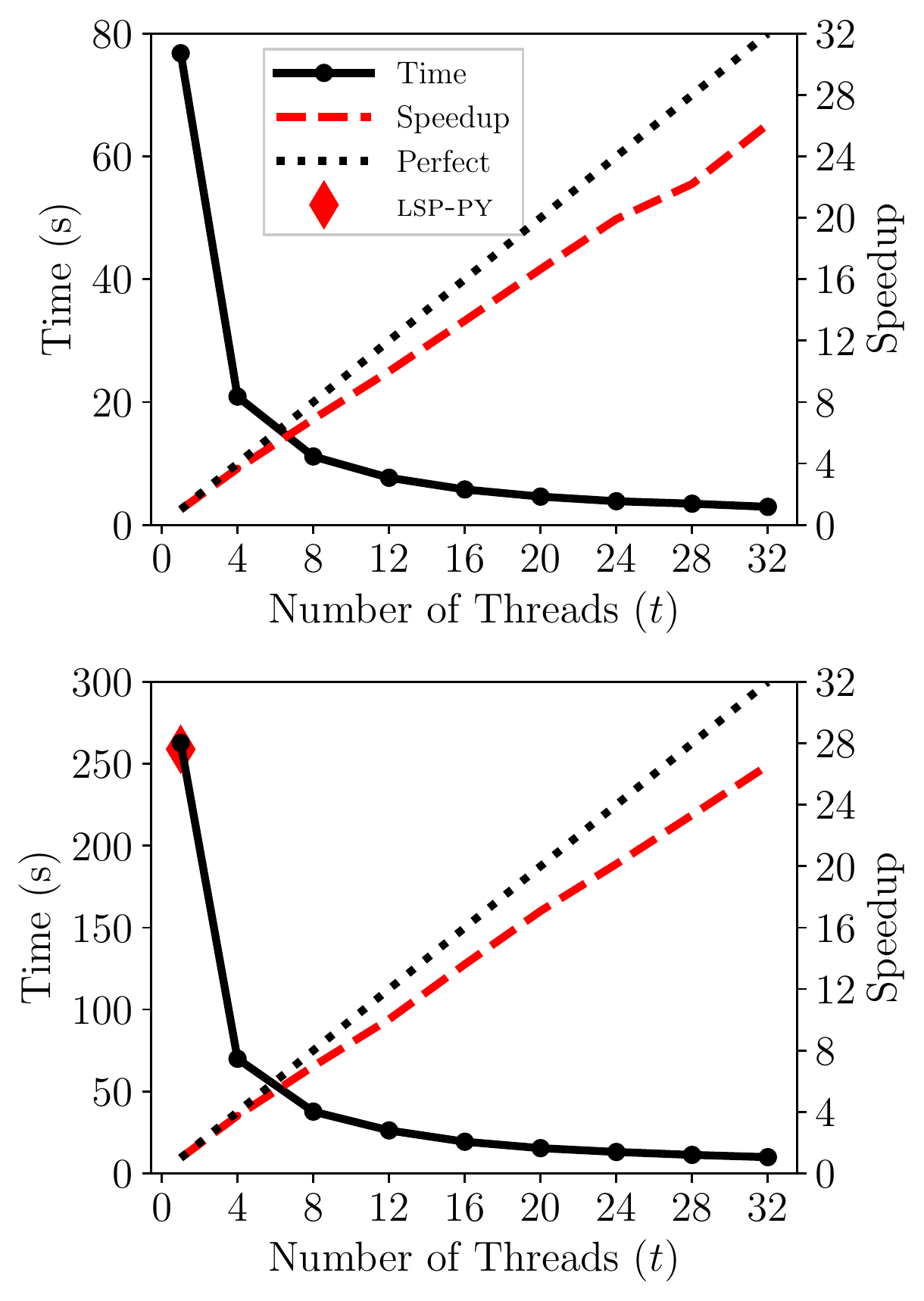}
	\caption{The scalability of \lspc. The response time is plotted on the left vertical axis as a function of the number of threads. On the right vertical axis, the speedup is shown. A perfect speedup is plotted for comparison. The experiment was executed on \platformb which has 32 total physical cores, using single object processing mode with $N_f=10^6$. The upper and lower panels correspond to FP32 and FP64 values, respectively.}
   \label{fig:scalability_cpu}
\end{figure}

We first assess the scalability of \lspc to understand the performance of the CPU algorithm. Figure~\ref{fig:scalability_cpu} plots the response time and speedup of \lspc as a function of the number of executing threads for FP32 (upper panel) and FP64 (lower panel) floating point values. We use single object mode using the dataset shown in Figure~\ref{fig:singleobj}. We select $N_f=10^6$ which is a large number of searched frequencies to demonstrate performance when there is a substantial amount of work to compute. For reference, a perfect speedup is plotted. Since the algorithm has a high compute to memory access ratio, the CPU cores have substantial work to compute and the algorithm achieves very good scalability. \lspc obtains a speedup (parallel efficiency) on $t=32$ cores of  26.09$\times$ (0.815) and 26.56$\times$ (0.830) on FP32 and FP64 values, respectively.

\begin{figure*}[!t]
\centering
  \includegraphics[width=0.8\textwidth]{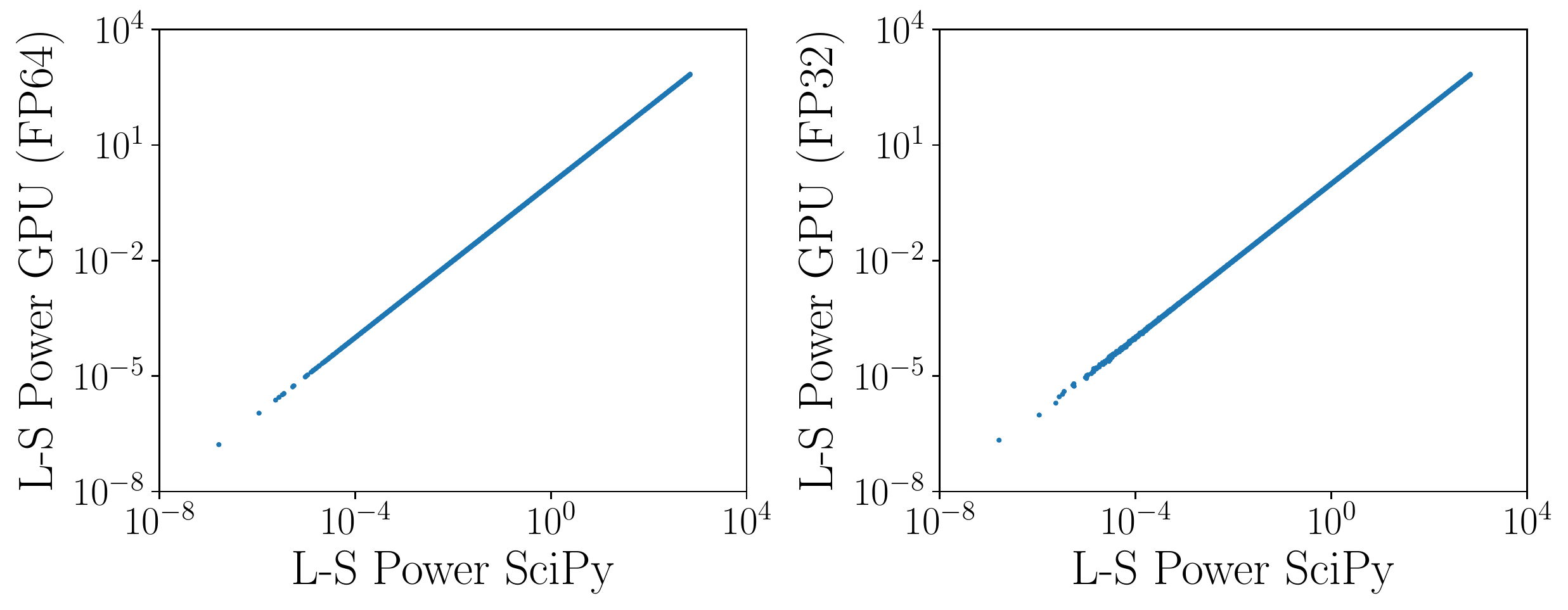}
	\caption{Comparison of periodograms generated by \gpu and \lsppy. The L-S power of \gpu as a function of L-S power of \lsppy is plotted for $N_f=10^6$ using single object mode. Execution of \gpu with FP64 and FP32 is shown on the left and right panels, respectively. The left panel shows near-perfect agreement between the periodograms generated by \lsppy and \gpu when FP64 is used. In the right panel, we observe that there is near-perfect agreement between \lsppy and \gpu when the power is sufficiently high ($\gtrsim 10^{-4}$). Therefore, executing L-S with only FP32 instead of FP64 is unlikely to lead to ambiguity when interpreting a periodogram. This also demonstrates that our GPU implementation is consistent with the output of the Python reference implementation.}
   \label{fig:single_object_accuracy_gpu_vs_scipy}
\end{figure*}

Using $t=32$ threads, we find that the ratio of the time to compute the LSP in double precision to single precision is 3.36, indicating that using double precision yields substantial performance degradation. Depending on the application,  when computing the LSP on the CPU, it may be preferable to use FP32 instead of FP64 due to this performance disparity. 

We plot the response time of the sequential SciPy LSP algorithm in Python (\lsppy) in the lower panel of Figure~\ref{fig:scalability_cpu}. We find that \lspc at $t=1$ and \lsppy have similar performance. This sanity check verifies that our ported code, \lspc, has consistent performance with the original implementation.

We omit showing scalability results for the batched execution mode, as results are similar.

\subsection{Accuracy}
In this section, we compare the accuracy of \gpu to \lsppy. Since \gpu can be executed using FP32 and FP64, we evaluate both options as compared to \lsppy which executes using FP64. Figure~\ref{fig:single_object_accuracy_gpu_vs_scipy} plots the L-S periodogram power of \lsppy vs. \gpu for FP64 (left panel) and FP32 (right panel), where a diagonal line indicates agreement between both methods. We execute both L-S algorithms using single batch mode using a frequency grid with $N_f=10^6$. We find good overall agreement between \lsppy and \gpu, but we observe that for FP32, some of the powers do not fall directly on the line at the lower end of the power range.

\begin{figure*}[!t]
\centering
  \includegraphics[width=0.8\textwidth]{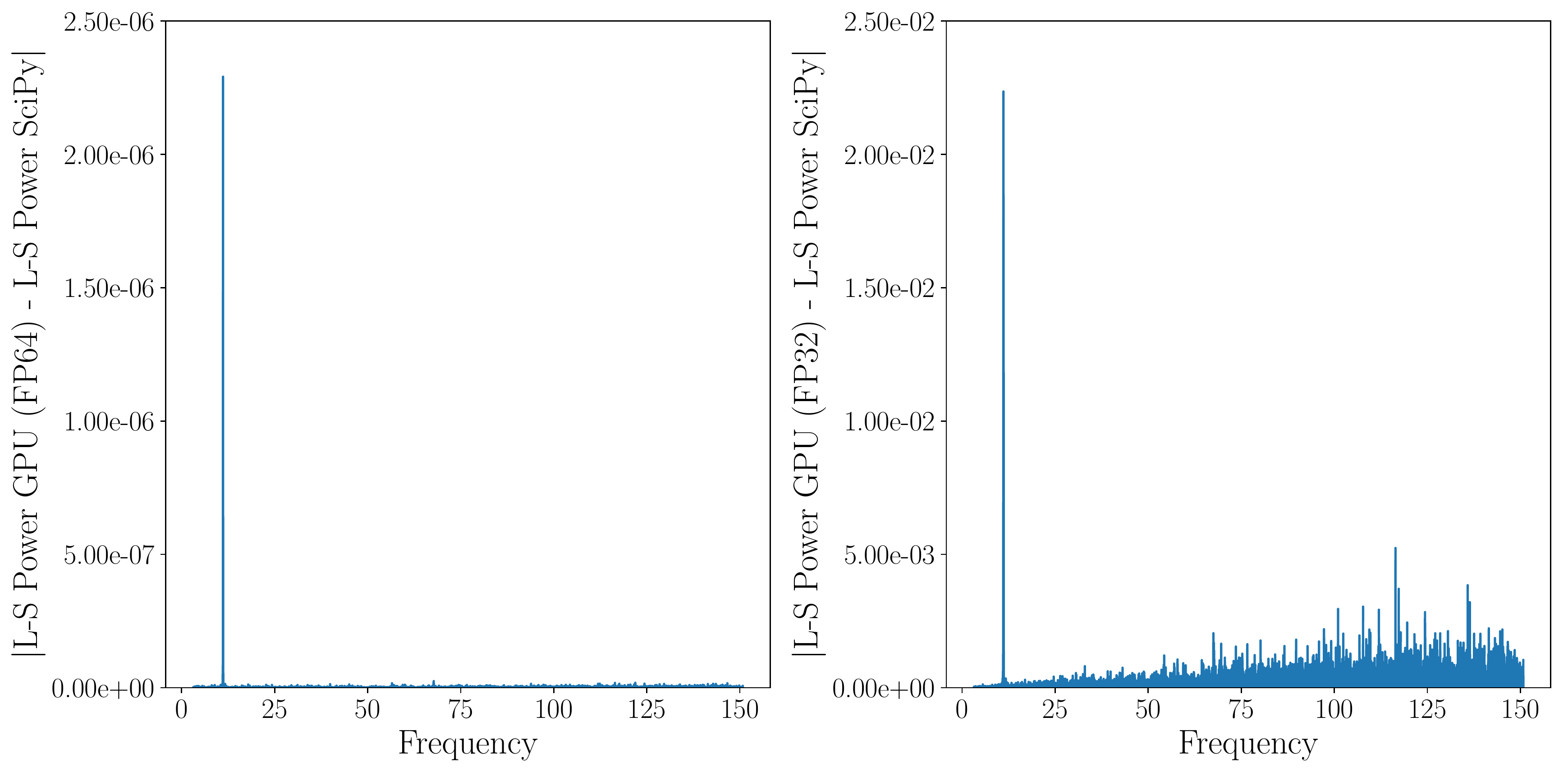}
	\caption{The absolute difference between the periodograms generated by \gpu and \lsppy using the data in Figure~\ref{fig:single_object_accuracy_gpu_vs_scipy}. Execution of \gpu with FP64 and FP32 is shown in the left and right panels, respectively.}
   \label{fig:single_object_accuracy_absolute_difference_pgram}
\end{figure*}

Figure~\ref{fig:single_object_accuracy_absolute_difference_pgram} plots the absolute difference between the periodograms using the same data shown in Figure~\ref{fig:single_object_accuracy_gpu_vs_scipy}. We find that \gpu executed with FP64 is consistent with \lsppy, with the highest error occurring around the frequency with the greatest power. The relative error is on the order of $10^{-6}$, which is negligible. Comparing \lsppy to \gpu executed with FP32, we find that the error is much higher than FP64, reaching a relative error on the order of $10^{-2}$. Despite this error, it is still low and unlikely to cause ambiguity when interpreting a periodogram. Overall, these results indicate that: $(i)$ the FP32 functionality of \gpu is likely sufficient for many applications, and $(ii)$ the results of our FP32 and FP64 \gpu implementations are consistent with the Python reference implementation.

\subsection{Transferring the Periodogram to the Host}
Data transfers over PCIe are a bottleneck because the bandwidth is lower than that between the CPU and main memory. Therefore, GPU applications that are bound by PCIe data transfers may not perform well compared to a parallel CPU implementation. In Section~\ref{sec:data_transfer_host}, we outlined two methods for transferring the data from the GPU to the host using either a standard \texttt{cudaMemcpy} or using several pinned memory staging buffers and transferring the data in three CUDA streams using \texttt{cudaMemcpyAsync}. Since the standard approach requires the driver to generate temporary pinned memory buffers for data transfers, it may be preferable to have the programmer manage the memory manually by making their own pinned memory buffers that are reused. 

In this experiment, we present the speedup of the pinned memory approach over the standard data transfer approach showing the total end-to-end response time, where \gpu uses batch mode and the global memory kernel. We elect to use batch mode as it requires a significant amount of data to be transferred from the GPU to the host, where the space complexity is $O(N_fN_o)$. As we will show in Section~\ref{sec:eval_SM}, the global and shared memory kernels achieve similar performance, so it does not matter which kernel we select here. Additionally, we examine both FP32 and FP64 data types since the latter requires double the amount of data to be transferred. 

\begin{figure}[!t]
\centering
  \includegraphics[width=1\columnwidth]{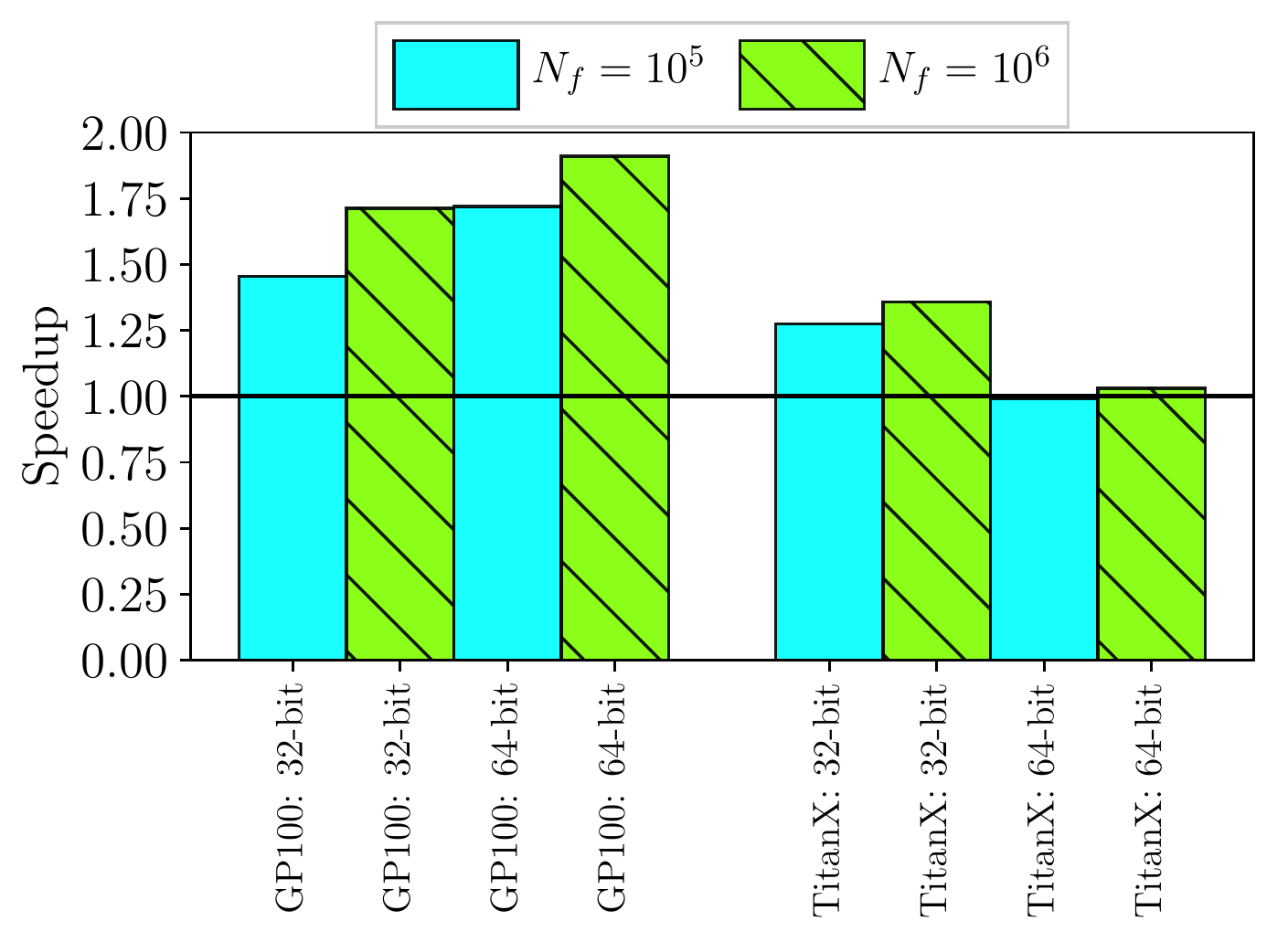}
	\caption{Speedup of using pinned memory data transfers in CUDA streams over standard memory copies for $N_f\in\{10^5, 10^6\}$ using batch mode. Executions are performed on both \platforma and \platformb for FP32 and FP64 floating point values using the global memory kernel. Values over the horizontal line indicate a performance gain from using the pinned memory data transfer scheme.}
   \label{fig:pinned_memory_data_transfer}
\end{figure}

Figure~\ref{fig:pinned_memory_data_transfer} plots the speedup of using the pinned memory approach vs. the standard approach on the GP100 and TitanX GPUs. On the GP100 GPU, we obtain a speedup by optimizing data transfers of up to 1.91$\times$. Computing on FP32 and FP64 yields a performance gain when using pinned memory. Despite FP64 values requiring more data to be transferred than FP32 values, the kernel execution time scales with the data transfer time, and so we observe performance gains on both FP32 and FP64.   
On the TitanX GPU, we obtain a speedup when processing FP32 data (up to 1.36$\times$); however, we do not observe a speedup on FP64 data. Since the TitanX GPU has fewer resources dedicated to FP64 arithmetic than the GP100, the execution time using FP64 values is bounded by kernel computation, and the fraction of time spent performing data transfers is negligible. This demonstrates that on the TitanX using FP32, and FP32 and FP64 on the GP100, the fraction of time performing device-to-host data transfers is non-negligible. In other words, the GPU is so efficient at computing the LSP that we observe data transfers requiring non-negligible time.

\subsection{The Impact of Using Shared Memory in the GPU Kernel}\label{sec:eval_SM}
As discussed in Section~\ref{sec:SM}, we compare the performance of two kernel designs. For each frequency, all input data values (time and magnitude) need to be read from global memory. We proposed two options to perform reading the input values from global memory, where the first option is to read the values directly from global memory, and the second option is to tile the computation, where the threads first page the data values into shared memory (each thread copies one element), and then all threads iterate over the data. Table~\ref{tab:global_vs_shared_memory} compares the response time of the batch and single object mode kernel execution times for the shared and global memory kernels, where the speedup of the shared memory over the global memory kernel is shown. Since the optimization only applies to the kernel, we present the kernel execution time and not the total end-to-end computation time.  We find that the shared memory kernel improves performance on all values of $N_f$ examined, achieving a speedup of up to 1.30$\times$ using batched mode with $N_f=10^4$. While this performance gain may seem significant, the speedup is calculated using the kernel execution time not the end-to-end time. The speedup using the end-to-end time is negligible in most cases. Therefore, since data transfers and other overheads have a major performance impact, improving the performance of the kernel does not contribute significantly to improving the end-to-end response time.  

While exploiting shared memory has been of critical importance on older generations of GPUs, newer generations have high global memory throughput and new cache technology; therefore, while shared memory improves the performance of \gpu, the performance gain is marginal in some cases. Additionally, the performance of the shared memory kernel is also a function of the selected CUDA block size and the length of the time series of each object ($N_t$). Consequently, it is possible that another configuration may degrade the performance of the shared memory kernel over the global memory kernel. Thus, in all that follows, we use the global memory kernel, despite the minor performance advantages offered by the shared memory kernel shown here. Overall, the importance of using shared memory has decreased over time and this is documented in the Nvidia marketing materials describing each successive generation of GPUs.

\begin{table*}[!t]
\centering
\footnotesize
\caption{\gpu kernel execution time (s) comparing global memory and shared memory kernels for $N_f\in \{10^4, 10^5, 10^6\}$. The speedup of the shared memory kernel over the global memory kernel and the speedup using the end-to-end response times are shown. \gpu is executed on \platforma (GP100) using FP64 on batch and single object processing modes. The response times are shown to five decimal places and the speedup is shown to two decimal places.}
\label{tab:global_vs_shared_memory}
\begin{tabular}{lrrrrrrrr} 

\hline
\multirow{2}{*}{$N_f$}&\multicolumn{4}{c}{Batch Mode}&\multicolumn{4}{c}{Single Object Mode}\\
\multicolumn{1}{l}{}&Global&Shared&Speedup (Kernel)&Speedup (Total) & Global&Shared&Speedup (Kernel)&Speedup (Total)\\\hline
$10^4$&0.03949& 0.03033& 1.30&1.04&0.00179& 0.00171& 1.05&1.01\\
$10^5$&0.34451& 0.32100& 1.07&1.09&0.00686& 0.00649& 1.06&1.05\\
$10^6$&3.19483& 3.03215& 1.05&1.02&0.05751& 0.05607& 1.03&1.00\\

\hline
\end{tabular}
\end{table*}

\subsection{Performance Evaluation of \gpu and \lspc}\label{sec:overview_best_configurations}
In the previous sections, we compared the merits and performance characteristics of two optimizations for \gpu. In this section, we configure \gpu to return the periodogram to the host using pinned memory data transfers and use the global memory kernel.

The LSP algorithm scales as $O(N_tN_f)$; therefore, to understand the performance of the algorithm across varying workloads, we can vary either the $N_f$ or $N_t$ parameter. Since our GPU algorithms are parallelized across frequencies, we elect to examine the performance across values of $N_f$, as performance may be more sensitive to $N_f$ than $N_t$.

\begin{figure}[!t]
\centering
  \includegraphics[width=1\columnwidth]{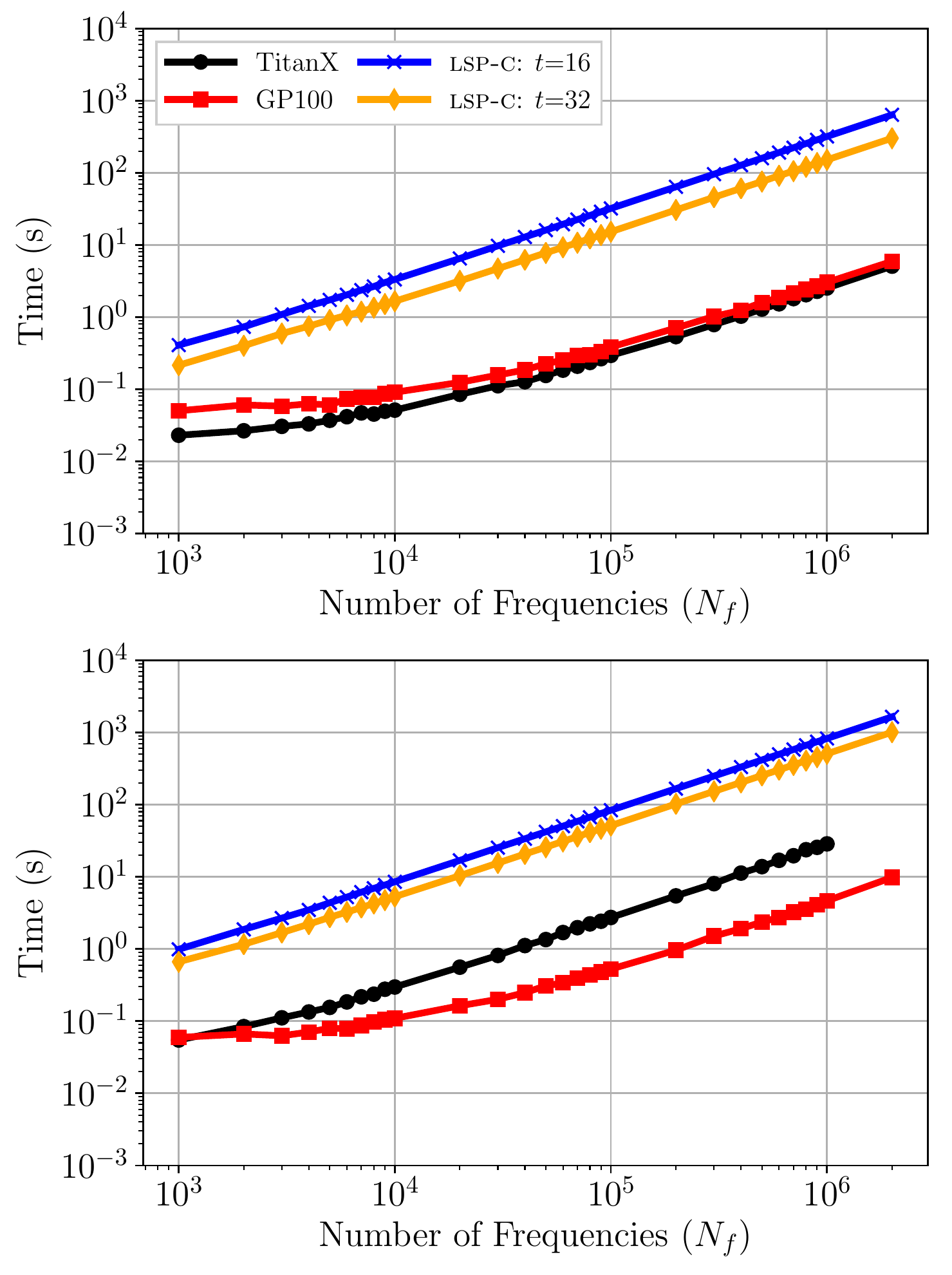}
	\caption{Total response time (s) as a function of $N_f$ comparing \gpu to \lspc using the batch mode. \gpu is configured to return the periodogram, uses pinned memory to transfer the data, and uses the global memory kernel. \lspc is configured with $t=16$ and $t=32$ threads when executed on \platforma and \platformb, respectively. The upper and lower panels show FP32 and FP64 floating point values, respectively.}
   \label{fig:time_vs_frequency_batch}
\end{figure}

Figure~\ref{fig:time_vs_frequency_batch} plots the response time as a function of $N_f$ for the batch mode comparing \gpu and \lspc. We plot results for both platforms illustrating performance on the GP100, TitanX, and the CPUs equipped with $t=16$ and $t=32$ threads on \platforma and \platformb, respectively (the number of physical cores on these platforms). In the top panel of Figure~\ref{fig:time_vs_frequency_batch} that reports the results for FP32 floating point values, we find that the response time of \gpu flattens at $\lesssim2\times 10^4$ on the GP100 and $\lesssim5\times 10^3$ on the TitanX. This is because these workloads are small and there may not be enough work to saturate GPU resources or fully amortize GPU overheads. Since the workload is higher on FP64 values, this effect is less pronounced in the lower panel. In the upper panel, we find that the TitanX outperforms the GP100 across all values of $N_f$. Since the TitanX represents the consumer-grade segment of the GPU market, and the GP100 represents the scientific computing/data center market, this result may seem surprising. However, since the TitanX is designed for FP32 operations, its FP32 capabilities are very similar to the GP100.  The lower panel of the figure (FP64 values)  demonstrates the benefit of the GP100 over the TitanX, where we observe that the GP100 achieves a speedup up to 6.66$\times$ over the TitanX. In the lower panel, we note the missing data point at $N_f=2\times10^6$ on the TitanX curve where there was insufficient global memory to store the periodogram data.

Unlike the performance of the GPU algorithms, \lspc performance in Figure~\ref{fig:time_vs_frequency_batch} does not flatten with small values of $N_f$. This is because the CPU has two orders of magnitude fewer cores, so they are saturated work work at all values of $N_f$. In contrast, the GPU typically needs to execute a factor of a few more threads than cores to hide global memory access latency and to take advantage of the high memory bandwidth on the device\footnote{\url{https://docs.nvidia.com/cuda/cuda-c-best-practices-guide/index.html}}.

Comparing \gpu to \lspc in Figure~\ref{fig:time_vs_frequency_batch}, we find that the TitanX achieves a maximum speedup over \lspc with $t=32$ of 59.69$\times$ and 19.15$\times$ on FP32 and FP64 floating point values, respectively. Likewise the GP100 achieves a maximum speedup over \lspc with $t=32$ of 50.95$\times$ (FP32) and 113.87$\times$ (FP64). The speedups are even greater on \platforma with $t=16$ cores. The performance gain of using the GPU over the CPU is staggering. However, since the LSP algorithm performs a significant amount of computation, has few branch conditions, and has regular memory access patterns, it is an ideal algorithm to execute on GPU hardware.

\begin{figure}[!t]
\centering
  \includegraphics[width=1\columnwidth]{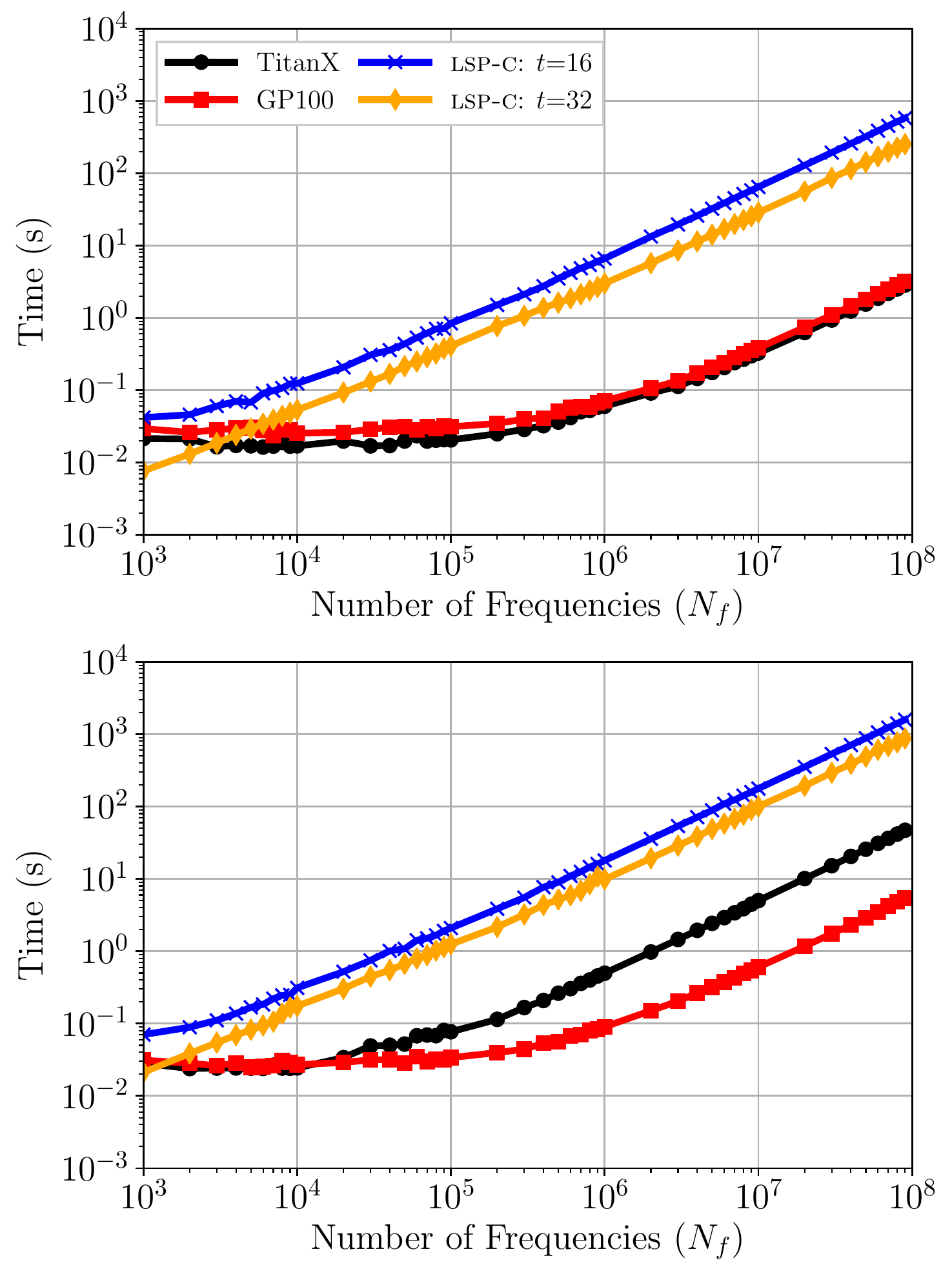}
	\caption{The same as Figure~\ref{fig:time_vs_frequency_batch}, but using single object mode.}
   \label{fig:time_vs_frequency_singleobj}
\end{figure}

Figure~\ref{fig:time_vs_frequency_singleobj} plots the same information as Figure~\ref{fig:time_vs_frequency_batch}, but we use single object mode. In this mode we only need storage space for one periodogram (instead of $N_o$ periodograms in batch mode), so we extend the range of $N_f$ values. As discussed by \citet{2018ApJS..236...16V}, the LSST may require up to 25 million frequency evaluations per object; therefore, we show $N_f\lesssim 10^8$.  From Figure~\ref{fig:time_vs_frequency_singleobj}, we observe similar performance behavior as in Figure~\ref{fig:time_vs_frequency_batch}. Particularly, the GPU remains undersaturated with work at lower values of $N_f$, which explains the flat response time at $N_f\lesssim10^5$.

Table~\ref{tab:32_vs_64_bit_floats} presents the maximum ratio of the FP64 to the FP32 response times across all values of $N_f$ in Figures~\ref{fig:time_vs_frequency_batch}~and~\ref{fig:time_vs_frequency_singleobj} for each algorithm on both platforms. We select the maximum ratio as it represents the worst case performance degradation due to using FP64 over FP32.  Comparing the GP100 and the TitanX on single object mode, we observe that the performance degradation of using FP64 floats requires a factor 1.69 more time for the GP100, whereas this factor is 16.62 on the TitanX. As discussed previously, the TitanX has few resources dedicated to FP64 arithmetic which causes this large performance disparity between processing FP32 and FP64 values. Interestingly, the performance penalty for using FP64 values on \lspc is higher than it is for the GP100. Comparing \lspc on both platforms, we observe that the ratio is higher when executing with $t=32$ than $t=16$, which we would not expect, as the number of cycles to perform FP32 and FP64 should be the same on both platforms. This difference is likely due to other factors that occur when executing the algorithm in parallel with more threads, such as an increased number of cache misses due to threads competing for space in higher levels of the memory hierarchy. As expected, we observe similar performance ratios when using batch mode so we omit interpreting these results.

\begin{table}[!t]
\centering
\footnotesize
\caption{Maximum response time ratio of the FP64 to FP32 values in Figures~\ref{fig:time_vs_frequency_batch}~and~\ref{fig:time_vs_frequency_singleobj}, corresponding to batch mode and single object mode, respectively.}
\label{tab:32_vs_64_bit_floats}
\begin{tabular}{lrrrr} 
\hline
~&GP100&TitanX&\lspc $t=16$&\lspc $t=32$\\\hline
Batch        &1.66&11.61&2.62&3.39\\
Single Object&1.69&16.62&2.82&4.00\\
\hline
\end{tabular}
\end{table}

Table~\ref{tab:speedup_summary} summarizes the speedup of the GPU over the CPU algorithms as executed on both platforms using the data in Figures~\ref{fig:time_vs_frequency_batch}~and~\ref{fig:time_vs_frequency_singleobj}. We report the maximum speedup obtained by the GPU implementation, but note that the CPU is competitive with the GPU at low values of $N_f$.  However, as discussed above, these small values of $N_f$ are likely unsuitable in practice,  as they may miss the peaks in the periodogram.




\begin{table*}[!t]
\centering
\footnotesize
\caption{Summary of the maximum speedup of \gpu over \lspc for the batched and single object modes from Figures~\ref{fig:time_vs_frequency_batch}~and~\ref{fig:time_vs_frequency_singleobj}.}
\label{tab:speedup_summary}
\begin{tabular}{lrrrrrrrr} 

\hline
&\multicolumn{4}{c}{Batch Mode}&\multicolumn{4}{c}{Single Object Mode}\\\hline
&GP100 FP32&GP100 FP64&TitanX FP32&TitanX FP64&GP100 FP32&GP100 FP64&TitanX FP32&TitanX FP64\\
$t=16$&107.31&186.43&125.64&31.18&181.00&306.07&207.09&37.57\\
$t=32$& 50.95&113.87& 59.69&19.15&79.25&174.53&92.24&23.50\\\hline

\hline
\end{tabular}
\end{table*}

\subsection{Generalized Perioogram with Error}\label{sec:generalized_LSP}
We have examined the performance of the standard LSP algorithm. However, as discussed by \citet{2011ApJ...733...10R}, the floating mean method in the generalized algorithm can allow for more robust period searches when the phase is sampled unevenly. We directly implement the Astropy geneneralized LSP algorithm that takes as input the photometric error and floats the mean\footnote{\url{https://github.com/astropy/astropy/blob/master/astropy/timeseries/periodograms/lombscargle/implementations/cython_impl.pyx}}. We implement both the batched and single object modes using the same configurations outlined in Section~\ref{sec:overview_best_configurations}. On the GP100, using FP64 with $n_f=10^6$ searched frequencies, the generalized periodogram requires a factor 2.06 and 2.14 of the response time of the standard algorithm on the batched and single object modes, respectively.


\subsection{Comparison to Townsend (2010)}
As discussed in Section~\ref{sec:intro}, this paper builds on the work of \citet{2010ApJS..191..247T}, as we have added functionality to the GPU-accelerated L-S algorithm. Here, we make a performance comparison to the GPU algorithm in \citet{2010ApJS..191..247T} denoted as \culsp. 

The \culsp algorithm as outlined in \citet{2010ApJS..191..247T} and the code located on the author's website uses FP32\footnote{\url{http://www.astro.wisc.edu/~townsend/resource/download/code/culsp.tar.gz}}. To make a comparison to our work, we modified \culsp to use FP32 or FP64. Furthermore, \culsp uses an intrinsic function \sincosfintrinsic which implements fast sin and cos functions in hardware at the expense of accuracy. The \sincosfintrinsic intrinsic is only available for FP32 and not FP64. Therefore, for comparison purposes, when executing \culsp, we either enable or disable \sincosfintrinsic; when disabled, we use the corresponding library function. In our implementation, \gpu, we do not use \sincosfintrinsic as it is unavailable for FP64. For code maintainability purposes, we elect to only use the library function (we may update the code if the intrinsic is available for FP32 and FP64 in the future). 

To make a fair comparison between approaches, we compare \culsp to \gpu with the shared memory kernel as both kernels use shared memory for paging the time series from global memory.    Furthermore, as we have shown throughout this paper, data transfers require a significant amount of time (Figure~\ref{fig:pinned_memory_data_transfer}). Since \culsp does not have batch modes or data transfer optimizations, we only compare kernel execution times using the single object mode.

Table~\ref{tab:comparison_townsend} shows the \culsp and \gpu kernel execution times for $N_f\in \{10^4, 10^5, 10^6\}$ using single object mode. As described above, \gpu does not employ the \sincosfintrinsic function, which is denoted by N/A in that column.   Comparing \culsp executed with FP32 with and without \sincosfintrinsic, we observe significant performance gains when using the intrinsic, which was also reported by \citet{2010ApJS..191..247T}. Comparing \gpu to \culsp without the intrinsic functions, we find that \gpu achieves an appreciable speedup over \culsp on both FP32 and FP64.  

We add several caveats to this analysis. \citet{2010ApJS..191..247T} proposed \culsp in 2010 before the development of several architectural advancements made in GPU hardware. This paper updates that pioneering work by including several additional optimizations and features. While we attempted to make a fair comparison between approaches, \culsp could potentially be updated to achieve better performance gains. Because both \gpu and \culsp use similar formulations of the L-S algorithm, there are few algorithmic differences. Both algorithms have several commonalities, such as using the \sincos function and using shared memory (however, we showed that shared memory is no longer advantageous over a standard global memory kernel). Given the similarities between algorithms, we do not expect large performance gains over \culsp. To reiterate, the main difference between this paper and \citet{2010ApJS..191..247T} is that we have included additional  functionality and optimizations that were unavailable in CUDA when \culsp was developed.

\begin{table}[!t]
\centering
\footnotesize
\caption{Kernel execution time (ms) comparing the shared memory \gpu kernel to \culsp, which also uses shared memory. We examine $N_f\in \{10^4, 10^5, 10^6\}$ on the single object dataset. Experiments are performed on \platforma. Intrinsic refers to using the \sincosfintrinsic intrinsic function.}
\label{tab:comparison_townsend}
\begin{tabular}{Xrrrrrrr} 
\hline
\multirow{1}{*}{$N_f$}&\multicolumn{2}{c}{FP32- Intrinsic}&\multicolumn{2}{c}{FP32- No Intrinsic}&\multicolumn{2}{c}{FP64}\\\hline
&\gpu&\culsp&\gpu&\culsp&\gpu&\culsp\\
$10^4$&N/A&0.151& 0.983&  1.486&1.602 & 1.676\\
$10^5$&N/A&1.088& 3.945&  5.531&6.396 & 9.860\\
$10^6$&N/A&9.821&35.469& 58.135&55.953&84.441\\
\hline
\end{tabular}
\end{table}

\subsection{Discussion}
\citet{2010ApJS..191..247T} reported a that their GPU algorithm achieved a speedup of $27.88\times$ over their parallel CPU algorithm on FP32 floating point values. We report a speedup of up to 181$\times$ on FP32 and 306$\times$ on FP64 data when evaluating a single object, which indicates that the performance disparity between the CPU and GPU has increased significantly over the past decade. \citet{2013MNRAS.434.3423G} compared period finding algorithm accuracy and performance and ported the CUDA code of \citet{2010ApJS..191..247T} to OpenCL. To our knowledge, these are the only publications proposing GPU-accelerated LSP algorithms.

In addition to deriving periods for Solar System objects from LSST data, there are many other applications that can use the LSP algorithm. One application is searching for stellar rotation periods in light curves from the Transiting Exoplanet Survey Satellite \citep{martins2020search}, and another is searching for the orbital periods of exoplanets \citep{2009A&A...496..577Z}. Period searches are computationally expensive, which can reduce the throughput of data processing pipelines that classify variable stars \citep{2011ApJ...733...10R}. \gpu can be utilized in these applications and others to reduce the computational burden of examining large scale time series datasets. Since multiple objects need to be examined, such as in those applications described above, the batch mode feature of our software can be utilized for this task.       

Several algorithmic advancements have been made to the LSP algorithm that reduce the $O(N_t^2)$ time complexity. Other works have proposed  $O(N_t\mathrm{log}N_t)$ variants of the LSP algorithm \citep{1989ApJ...338..277P,2012A&A...545A..50L}. In particular, \citet{2012A&A...545A..50L} compared the performance of their $O(N_t\mathrm{log}N_t)$ algorithm on the CPU to the $O(N_t^2)$ GPU algorithm proposed by \citet{2010ApJS..191..247T}, and found that the execution time of their algorithm is roughly 5 times shorter than the GPU algorithm and obtains high accuracy relative to the na\"{i}ve algorithm.  $O(N_t\mathrm{log}N_t)$ algorithms are another approach to reduce the execution time of the LSP algorithm; however, given current hardware trends, it may be preferable to perform the na\"{i}ve LSP algorithm on the GPU to achieve good performance while avoiding potential accuracy loss.

\section{Conclusion}\label{sec:conclusion}
The primary motivation for this work is the use of LSP to compute in near real-time the rotation periods for batches of asteroids that our team will receive from the LSST. Using our batched mode on $\approx 1,000$ synthetic asteroids, and a reasonable value of $N_f=2\times10^5$, we are able to compute the periods on the GPU and return the periodogram in $\approx 1$~s using FP64 precision. This implies that we will have a leftover time budget of $\sim29$~s to perform outlier detection activities and send alerts to the Solar System community  before the next batch of asteroids arrives in the LSST data stream. 

We have demonstrated that \gpu yields superior performance over the parallel CPU implementation. Depending on the application, it may be preferable to use FP32 instead of FP64 floating point values to avoid the cost associated with higher precision or to use a GPU that has minimal resources dedicated to 64-bit arithmetic. On a pragmatic note, it is likely that FP32 arithmetic is sufficient for the LSP algorithm, but our software allows the user to select the level of precision that they believe to be appropriate for their scientific investigation. An overview of the code versions is given in~\ref{sec:code}.


Future work includes investigating other period finding algorithms on the GPU, such as algorithms that assume more structure in the data compared to the L-S algorithm, such as Super Smoother \citep{friedman1984variable}.

\section*{Acknowledgment}
This work has been supported in part by the Arizona Board of Regents, Regents’ Innovation Fund.
We thank the anonymous reviewer for their insightful comments and helpful feedback on our manuscript.

\appendix
\section{Open Source Code}\label{sec:code}
The source code is publicly available at~\url{https://github.com/mgowanlock/gpu_lomb_scargle}. To summarize, our code assumes the following: $(i)$ An evenly spaced frequency grid is used, defined by minimum and maximum frequency ranges and the number of searched frequencies. $(ii)$ The algorithm uses angular frequencies. The frequency is given by $f=2\pi(p^{-1})$, where $p$ is the period. $(iii)$ The units output by the code are given by the units in the input dataset file. For clarity, in the paper, we converted Julian Date (days) to hours.

The code has two versions. One version contains the code used to produce the results in this paper, including the two GPU and CPU implementations for the single object and batched processing modes. To reduce confusion with all of the parameters used in the paper, we also include another version that has a default set of parameters selected for the user. In particular, this version of the code uses the global memory kernel, uses the pinned memory optimization for data transfers, and returns the periodogram(s) to the host. The user must input the frequency ranges, number of frequencies to search, the enumerated data type (FP32 or FP64), whether the period(s) for the object(s) should be printed to standard out, and select the standard or generalized periodogram mode. This version also automatically detects whether the user wants to compute the LSP of a batch of objects or a single object.

\section*{References}


\end{document}